\pdfoutput=1

\documentclass[twocolumn,showpacs]{revtex4}
\usepackage{graphicx,amssymb,amsmath}
\usepackage[usenames,dvipsnames]{color}    
\begin{document}

\title{
  Pseudodiffusive conductance, quantum-limited shot noise, and Landau-level hierarchy in biased graphene bilayer
}

\author{Grzegorz Rut}
\affiliation{Marian Smoluchowski Institute of Physics, 
Jagiellonian University, Reymonta 4, PL--30059 Krak\'{o}w, Poland}
\author{Adam Rycerz}
\affiliation{Marian Smoluchowski Institute of Physics, 
Jagiellonian University, Reymonta 4, PL--30059 Krak\'{o}w, Poland}

\begin{abstract}
We discuss, by means of mode-matching analysis for the Dirac equation, how splittings of the Landau-level (LL) degeneracies associated with spin, valley, and layer degrees of freedom affect the ballistic conductance of graphene bilayer. The results show that for wide samples ($W\gg{}L$) the Landauer-B\"{u}ttiker conductance reaches the maximum $G\simeq{}se^2/(\pi{h})\times{}W/L$ at the resonance via each LL, with the prefactor varying from $s=8$ if all three degeneracies are preserved, to $s=1$ if all the degeneracies are split. In the absence of bias between the layers, the degeneracies associated with spin and layer degrees of freedom may be split by manipulating the doping and magnetic field; the conductance at the zeroth LL is twice as large, while the conductance at any other LL equals to the corresponding conductance of graphene monolayer. The presence of bias potential allows one also to split the valley degeneracy. Our results show that the charge transfer at each LL has pseudodiffusive character, with the second and third cumulant quantified by ${\cal F}=1/3$ and ${\cal R}=1/15$ (respectively). In case the electrochemical potential is allowed to slowly fluctuate in a~finite vicinity of LL, the resulting charge-transfer characteristics are still quantum-limited, with ${\cal F}\simeq{}0.7$ and ${\cal R}\simeq{}0.5$ in the limit of large fluctuations. Analogously, the above values of ${\cal F}$ and ${\cal R}$ are predicted to be approached in the limit of high source-drain voltage difference applied. The possible effects of indirect interlayer hopping integrals are also briefly discussed.
\end{abstract}

\date{\today}
\maketitle

\section{Introduction}

Several unique physical phenomena were observed in graphene or its derivatives at high magnetic fields \cite{Cas09,Das11,Goe11}. These include Shubnikov-de Haas oscillations indicating zero quasiparticle rest mass \cite{Nov05}, room-temperature quantum Hall effect with a nonstandard (half-odd integer) sequence of Landau levels \cite{Nov07}, signatures of a~fractal energy spectrum known as Hofstadter's butterfly \cite{Dea13}, and many others. This new subarea of condensed-matter physics emerges primarily due to the nature of effective quasiparticles, which are chiral Dirac fermions with zero (the case of graphene monolayer) or small effective masses ($m_{\rm eff}=0.033\,m_e$ in the case of graphene bilayer, with $m_e$ the free electron mass) coupled to the external electromagnetic field via additive terms in low-energy Hamiltonians, which are {\em linear} in both scalar and vector potentials \cite{linefoo}. A remarkable consequence of such a~coupling is the quantization of the visible light absorption \cite{Nai08}.

Among numerous phenomena which were predicted theoretically but not yet fully confirmed experimentally, we focus our attention on the so-called {\em pseudodiffusive} transport in ballistic graphene. For an~undoped monolayer, elementary mode-matching analysis for the Dirac equation \cite{Kat06a,Two06} leads to the Landauer-B\"{u}ttiker conductance \cite{Naz09} of a~rectangular sample (with the width $W$ and the length $L$) scaling as $G=\sigma_0\times{}W/L$ for $W\gg{}L$, where $\sigma_0=(4/\pi)e^2/h$ is the universal quantum value of the conductivity. Additionally, the Fano factor is ${\cal F}=1/3$, and all the other charge-transfer characteristics are indistinguishable from those of a classical diffusive conductor \cite{Ben08,Ryc09}. In the pseudodiffusive regime, applied magnetic field is predicted to affect neither the conductance \cite{Ost06,Lou07} nor other transport characteristics \cite{Pra07}. Existing experiments \cite{Mia07,Dic08,Dan08} generally support these theoretical results, leaving some ambiguity concerning the origin of the ${\cal F}$ value observed \cite{Das11,Wie11}. For high dopings and magnetic fields, charge transport through a~monolayer was discussed in analytical terms for the rectangular \cite{Pra07} and the disk-like (Corbino geometry) samples \cite{Che06,Ryc10}. In both cases, pseudodiffusive behavior is expected to be recovered at each resonance with the Landau level (LL) in the absence of disorder. Remarkably, recent numerical study of large disordered samples \cite{Ort13} reports the longitudinal conductivity $\sigma_{xx}\simeq{}1.4\,e^2/h$ (what is numerically close to $\sigma_0$) appearing at each LL for wide ranges of disorder and magnetic fields. The nature of this coincidence, however, remains unclear so far.

\begin{figure}
\centerline{\includegraphics[width=\linewidth]{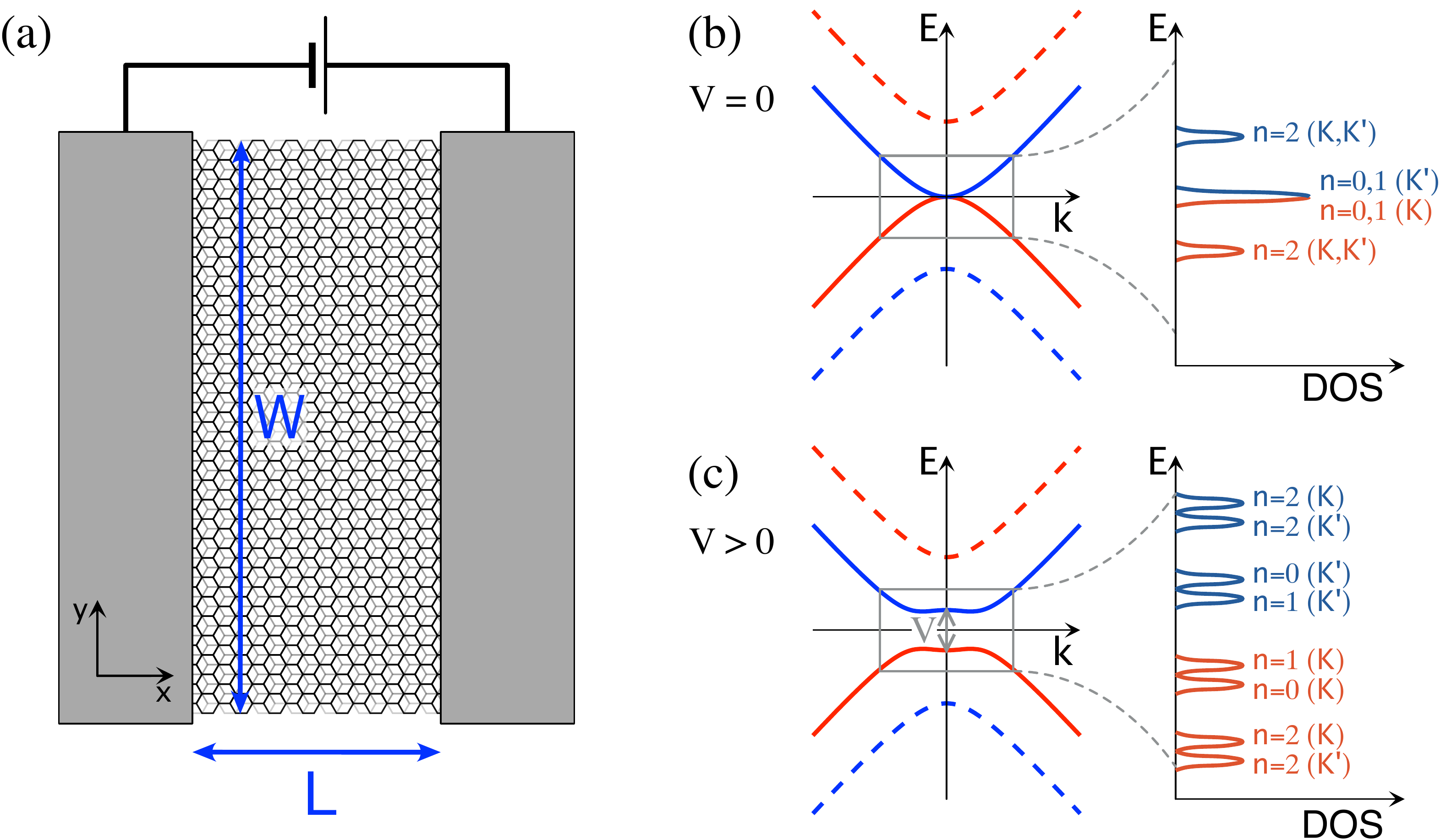}}
\caption{\label{setupfig}
  Schematics of system studied analytically in the paper and energy band structure in the quantum Hall regime. (a) A strip of graphene bilayer of width $W$ attached to two electrodes (shaded rectangles) at a~distance $L$. A voltage source drives a~current through the sample area. Separate top and bottom gate electrodes (not shown) allow one to tune the carrier concentration and the band gap (related to the potential energy difference between the layers $V$). (b,c) The formation of Landau levels in bilayer graphene with and without a band gap. Landau levels are indexed with the orbital index $n$ and the valley pseudospin ($K$ or $K'$); the twofold spin degeneracy of each level is assumed for clarity. In the absence of a band gap ($V=0$) almost every Landau level shows the fourfold (spin and valley) degeneracy, with the exception of eight-fold degenerate zero-energy level, for which the states arising from two layers (red and blue lines) coexist. Both layer and valley degeneracies are split in the presence of a band gap ($V>0$).
}
\end{figure}

For a~bilayer, a~few theoretical studies \cite{Kat06b,Cse07,Sny07} showed that regardless {\em massive} Dirac fermions govern low-energy properties of the system, the pseudodiffusive conductivity of undoped ballistic samples is $(8/\pi)e^2/h=2\sigma_0$ (twice as large as in the case of a~monolayer), and the Fano factor ${\cal F}=1/3$ again. Surprisingly, a role of the most desired property of graphene bilayer, which is a~tunability of the energy gap related to the potential energy difference between the layers $V$ \cite{Mac06,Cas07,Nil07,Per07}, has been only marginally discussed in the context of pseudodiffusive transport \cite{Fer11}. We notice here, that for a~Hall-bar setup (for which $W\lesssim{}L$ and the pseudodiffusive limit is usually inaccessible) it was shown both numerically and experimentally that the eightfold degeneracy of the lowest LL can be lifted by manipulating the external electromagnetic fields (see Fig.\ \ref{setupfig}), and the effect was usually attributed to electron-electron interactions \cite{Wei10,Lai08,Zha12}. 

Here, transport properties of graphene bilayer in the presence of potential energy difference between the layers and external magnetic fields are discussed in analytical terms. Namely, we start from the four-band Dirac Hamiltonian \cite{Mac06} taking into account the inter- and intralayer nearest neighbour hopping parameters, and employ the Landauer-B\"{u}ttiker formalism \cite{Naz09} to investigate the field-dependent conductance and other transport characteristics of a~ballistic sample. The geometry considered (wide-and-short sample) is chosen in such a~way that the boundary conditions applied to the Dirac equation do not affect the resulting physical quantities. 

The remaining part of the paper is organized as follows. In Section II we present the system details and find all linearly-independent solutions of the corresponding Dirac equation at finite dopings, biases, and magnetic fields. Then, in Section III we discuss the field-dependent transport characteristics in three different situations: at the Dirac point, in an unbiased sample ($V=0$) and in a sample with different potentials on the layers ($V\neq{}0$). In Section IV we analyze the influence of a~finite voltage difference or doping fluctuations (in the vicinity of pseudodiffusive regions), on the shot-noise power and on the third charge-transfer cumulant. Also in Section IV  we compare, with a~help of the so-called {\em partial conductance}, the statistical distribution of transmission probabilities for graphene bilayer in high magnetic fields with the corresponding distribution for a generic diffusive system. In Section V we discuss, by solving the appropriately modified Dirac equation numerically, the possible role of indirect interlayer hopping integrals. The conclusions are given in Section VI.

\section{The setup and mode-matching for the Dirac equation}

\subsection{The effective Hamiltonian}
Following Snyman and Beenakker \cite{Sny07}, we consider a rectangular, weakly doped bilayer sample attached to two heavily-doped strips modelling contacts [see Fig.\ \ref{setupfig}(a)]. It is also assumed that the magnetic field ($B\neq{}0$) is present only in the sample area. Our analysis starts from the four-band Hamiltonian for the $K$ valley \cite{Mac06}
\begin{equation}
  \label{eq:hamiltonian1}
  H=\left(\begin{array}{cccc}
      U_{1}(x) & \pi_{x}\!+\!i\pi_{y} & t_{\bot} & 0\\
      \pi_{x}\!-\!i\pi_{y} & U_{1}(x) & 0 & 0\\
      t_{\bot} & 0 & U_{2}(x) & \pi_{x}\!-\!i\pi_{y}\\
      0 & 0 & \pi_{x}\!+\!i\pi_{y} & U_{2}(x)
    \end{array}\right),
\end{equation}
where $t_{\perp}\simeq{}0.4\,$eV is the interlayer nearest-neighbour hopping energy, $\pi_j/v_{F}=\left(-i\hbar\,\partial_j+eA_j\right)$
is the gauge-invariant in-plane momentum operator ($j=1,2$), the electron charge is $-e$, and $v_{F}\simeq{}10^6\,$m/s is the Fermi velocity in a~single layer. $U_{l}(x)$ (with $l=1,2$ the layer index) is the electrostatic potential energy chosen as
\begin{equation}
  \label{eq:potential}
  U_{l}(x)=\begin{cases}
    U_{\infty} & \mbox{if \ensuremath{x<0\mbox{ or }x>L},}\\
    \lambda_{l}V-g\mu_{\rm B}B\,m_{s} & \mbox{if \ensuremath{0<x<L}},
  \end{cases}
\end{equation}
where $V$ is the difference between potentials on the layers, $\lambda_{l}=\frac{1}{2}(-1)^{l}$, and $g\mu_{\rm B}B m_{s}$ is the Zeeman term (the z-component of spin $m_{s}=\pm{}\frac{1}{2}$). The experimental values of the Lande factor for graphene bilayer are $g\simeq{}2-3$ \cite{Jia07,Kur11,Vol12}, thus we set $g=2$ for the numerical discussion. In order to obtain the Hamiltonian for the other valley ($K'$), it is sufficient to substitute $V\rightarrow{}-V$ and $\pi_j\rightarrow{}-\pi_j$ in Eq.\ (\ref{eq:hamiltonian1}).

\subsection{The sample area}
We choose the Landau gauge ${\bf A}\equiv{}(A_x,A_y)=(0,-Bx)$, with the uniform magnetic field $B\neq{}0$ for $0<x<L$ (otherwise, $B=0$). The wavefunction is a four-component spinor, which can be written as $\psi=\left(\phi_{A_{1}},i\phi_{B_{1}},\phi_{B_{2}},i\phi_{A_{2}}\right)^{T}$. The Hamiltonian (\ref{eq:hamiltonian1}) commutes with $-i\partial_y$, and thus $\psi$ varies in the y-direction as a plane wave of a form $\propto\exp\left(i{k_{y}}y\right)$, with the transverse wavenumber $k_y$. The Dirac equation for a~sample area, after a~substitution $\xi=l_{B}^{-1}x-k_{y}l_{B}$ (with $l_{B}=\sqrt{\hbar/|eB|}$ the magnetic length), can be written as
\begin{equation}
\label{eqfiafib}
  \left(\begin{array}{cccc}
      -\varepsilon-\delta & \partial_{\xi}+\xi & t & 0\\
      \partial_{\xi}-\xi & \varepsilon+\delta & 0 & 0\\
      t & 0 & -\varepsilon+\delta & \partial_{\xi}-\xi\\
      0 & 0 & \partial_{\xi}+\xi & \varepsilon-\delta
    \end{array}\right)\left(\begin{array}{c}
      \phi_{A_{1}}\\
      \phi_{B_{1}}\\
      \phi_{B_{2}}\\
      \phi_{A_{2}}
    \end{array}\right)=0,
\end{equation}
where we have defined $\varepsilon=\left(E-g\mu_{\rm B}B m_{s}\right)l_B/(\hbar{}v_{F})$, $\delta=-Vl_B/(2\hbar{}v_{F})$, and $t=t_{\perp}l_B/(\hbar{}v_{F})$. The functions $\phi_{\alpha}$ are given explicitly in Appendix~\ref{appfun}. Here we only mention that solutions at the Dirac point ($\varepsilon=\delta=0$) still have a~peculiar form of evanescent waves, leading to zero-field value of the pseudodiffusive conductance \cite{Sny07} unaltered for arbitrarily high magnetic fields. We address this issue in a~detailed manner in Section III.

\subsection{Contact regions}
For contact regions, one can neglect the bias potential ($\delta\simeq0$) due to a~high doping. The Dirac equation can thus be written as
\begin{equation}
\left(\begin{array}{cccc}
-\tilde{\epsilon} & ke^{i\theta_{k}} & \tilde{t} & 0\\
ke^{-i\theta_{k}} & -\tilde{\epsilon} & 0 & 0\\
\tilde{t} & 0 & -\tilde{\epsilon} & ke^{-i\theta_{k}}\\
0 & 0 & ke^{i\theta_{k}} & -\tilde{\epsilon}
\end{array}\right)\left(\begin{array}{c}
\phi_{A_{1}}\\
i\phi_{B_{1}}\\
\phi_{B_{2}}\\
i\phi_{A_{2}}
\end{array}\right)=0,
\end{equation}
with $\tilde{\epsilon}=(E-U_{\infty})/(\hbar\,\nu_{F}),\,\tilde{t}=t_{\perp}/(\hbar\,\nu_{F})$, $k=\sqrt{k_x^2+k_y^2}$, and $\theta_{k}=\mbox{arg}(k_{x}+i\, k_{y})$. After straightforward calculations, one obtains the dispersion relation
\begin{equation}
  \label{eq:dyspersja}
  \tilde{\epsilon}\left(k\right)^{2}=\left(\frac{\eta}{2}\,\tilde{t}+\sqrt{\frac{1}{4}\,\tilde{t}^{2}+k{}^{2}}\right)^{2},
\end{equation}
with $\eta=\pm{}1$ referring to the two subbands. 

The eigenfunctions in contact regions take the form of plane wave spinors, namely
\begin{align}
\label{psillim}
\psi_{L}^{\pm}\left(x\right) &= C\!\left(\tilde{\epsilon},k_{x}^{\pm}\right) \exp({-ixk_{x}^{\pm}})\left(\begin{array}{c}
\mp\tilde{\epsilon}\\
\pm\left(k_{x}^{\pm}+i\, k_{y}\right)\\
\tilde{\epsilon}\\
-k_{x}^{\pm}+i\, k_{y}
\end{array}\right), \\
\label{psirlim}
\psi_{R}^{\pm}\left(x\right) &= C\!\left(\tilde{\epsilon},k_{x}^{\pm}\right) \exp({ixk_{x}^{\pm}})\left(\begin{array}{c}
\mp\tilde{\epsilon}\\
\mp\left(k_{x}^{\pm}-i\, k_{y}\right)\\
\tilde{\epsilon}\\
k_{x}^{\pm}+i\, k_{y}
\end{array}\right).
\end{align}
For instance, one can model the heavily electron-doped contacts by taking the limit $U_{\infty}\rightarrow-\infty$, leading to $k_{x}^{\pm}=\sqrt{\tilde{\epsilon}\,\left(\tilde{\epsilon}\pm\tilde{t}\right)-k_{y}^{2}}\simeq{}|\tilde{\epsilon}|$.
The symbols $\psi_{R}^\pm$ and $\psi_{L}^\pm$ denote the solutions moving to the right and to the left (respectively), with the signs $\pm$ referring to the two subbands again. The normalization factors $C\!\left(\tilde{\epsilon},k_{x}^{\pm}\right)$ are chosen such that the total current $I_{L(R)}^{\pm}=ev_{F}\int_{0}^{W}dy\left(\psi_{L(R)}^{\pm}\right)^{\dagger}\left(\begin{array}{cc}
\sigma_{x} & 0\\
0 & \sigma_{x}
\end{array}\right)\psi_{L(R)}^\pm$ satisfies $|I_{L(R)}^{\pm}|=ev_F$, implying $C\!\left(\tilde{\epsilon},k_{x}^{\pm}\right)=1/\sqrt{4W\tilde{\epsilon} k_{x}^{\pm}}$.
 

\section{Transport of Dirac fermions}
In this Section we present our main results concerning the conductance $G$, the Fano factor ${\cal F}$, and the factor ${\cal R}$ quantifying the third charge-transfer cumulant for ballistic graphene bilayer. We employ the standard Landauer-B\"{u}ttiker formalism \cite{Naz09}, namely
\begin{eqnarray}
G & = & G_0\mbox{Tr}\,\boldsymbol{T},\label{eq:glan}\\
{\cal F} & = & \frac{\mbox{Tr}\left[\boldsymbol{T}\left(\mathbf{1}-\boldsymbol{T}\right)\right]}{\mbox{Tr}\,\boldsymbol{T}},\label{eq:ffac}\\
{\cal R} & = & \frac{\mbox{Tr}\left[\boldsymbol{T}\left(\mathbf{1}-\boldsymbol{T}\right)\left(\mathbf{1}-2\boldsymbol{T}\right)\right]}{\mbox{Tr}\,\boldsymbol{T}},\label{eq:rfac}
\end{eqnarray}
where $G_{0}=e^{2}/h$ is the conductance quantum, $\boldsymbol{T}=\boldsymbol{t^{\dagger}t}$, and $\boldsymbol{t}$ is a block-diagonal matrix with each block [of the form given by Eq.\ (\ref{t1block}) in Appendix~\ref{apptra}] corresponding to a~single transmission channel, identified by the valley index ($K$ or $K'$), the transverse momentum $k_y$, and the $z$-component of spin $m_s$. Details of the mode-matching analysis are given in Appendix~\ref{apptra}.

\subsection{Unbiased graphene bilayer \label{traunbigra}}
At zero doping and zero bias potential ($\varepsilon=\delta=0$) we obtain the transmission probabilities
\begin{equation}
T_{k_{y}}^{\pm}(0)=\mbox{cosh}^{-2}\left[\left(k_{y}-\frac{1}{2}l_{B}^{-2}L\pm k_{c}\right)L\,\right],\label{eq:trans0}
\end{equation}
where $k_{c}=\frac{1}{L}\mbox{ln}\left[\frac{Lt_{\perp}}{2\hbar{}v_{F}}+\sqrt{1+\left(\frac{Lt_{\perp}}{2\hbar{}v_{F}}\right)^{2}}\right]$ and the pairwise structure $\{T_{k_y}^+,T_{k_y}^-\}$ for a~given $k_y$ can be attributed to the presence of two graphene layers. In comparison to the case of bilayer graphene at the Dirac point at zero magnetic field studied in Ref.\ \cite{Sny07}, the wave vector is shifted by a factor $-l_{B}^{-2}L/2$, which is proportional to $B$. Provided the sample width is much larger than length ($W\gg{}L$) the boundary effects do not play an important role and one can choose the periodic boundary conditions; i.e., $k_y=2\pi{}n/W$ with $n=0,\pm{}1,\pm{}2,\dots$. In such a~limit, each of the sums over transverse momenta in Eqs.\ (\ref{eq:glan}--\ref{eq:rfac}) can be approximated by an integral according to
$$
  \sum_{k_y}\underset{\scriptscriptstyle W\gg{}L}{\simeq}
  W\int_{-\infty}^{\infty}\frac{dk_y}{2\pi}.
$$
In case the Zeeman splitting can be neglected ($g\simeq{}0$) this leads to the field-independent pseudodiffusive conductance twice as large as in the case of a~monolayer, i.e.\
\begin{equation}
  \label{gdiff12}
  G^{(2)}_{\rm diff}=2G_{\rm diff}^{(1)}=G_{0}\,\frac{8}{\pi}\,\frac{W}{L},
\end{equation}
where the upper index denotes the number of layers. Also, the shot-noise power and the third charge-transfer cumulant are field-independent and quantified by ${\cal F}\simeq{}1/3$ and ${\cal R}\simeq{}1/15$ (respectively).

\begin{figure}[!ht]
\centerline{
  \includegraphics[width=0.9\linewidth]{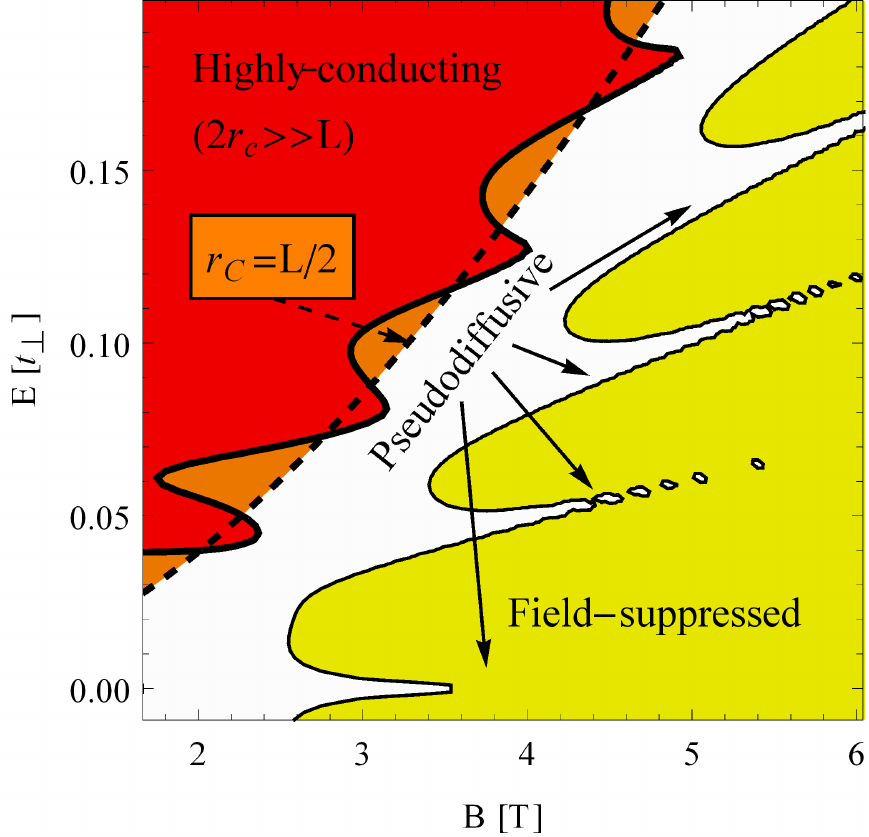}
}
\caption{\label{fig:Transport-regimes-in}
  Transport regimes in unbiased graphene bilayer (Zeeman splitting is not taken into account). Two solid lines delimit the areas with $G/G_{0}>8\,W/L$ (red) and $G/G_{0}<2.4\,W/L$ (yellow), where we set $W/L=20$ and $L=48\,\hbar{}v_F/t_\perp\simeq{}77\,$nm. Dashed line marks a~border of the highly-conducting regime following from Eq.\ (\ref{eq:rc}). 
}
\end{figure}

At finite dopings and zero bias potential ($\varepsilon\neq{}0$, $\delta=0$), one can identify three distinct transport regimes: the highly-conducting ($G\gg{}G^{(2)}_{\rm diff}$), the field suppressed ($G\ll{}G^{(2)}_{\rm diff}$), and the pseudodiffusive ($G\simeq{}G^{(2)}_{\rm diff}$), as depicted in Fig.\ \ref{fig:Transport-regimes-in}. The highly-conducting regime shows up in relatively weak fields, when the cyclotron radius $r_{C}=\hbar{}k/\left|eB\right|\gtrsim{}L/2$. Using the energy dispersion for the lower conductance (or the higher valence) subband given by Eq.\ (\ref{eq:dyspersja}), one can rewrite this condition as
\begin{equation}
  \label{eq:rc}
  \left|E\right|\gtrsim{}\frac{1}{2}\left[
    \sqrt{t_{\perp}^{2}+\left(\frac{\hbar{}v_F{}L}{l_B^2}\right)^{2}}-t_{\perp}
  \right].
\end{equation}
In stronger fields, the charge transport is suppressed and a~considerable conductance $G\gtrsim{}G_0$ emerges only in narrow energy intervals near LLs, in analogy with corresponding results for a~monolayer reported in Refs.\ \cite{Pra07,Ryc10}. For any of these intervals, it is possible to increase $B$ keeping the doping such that 
$
  \varepsilon^2\pm{}\sqrt{(\varepsilon{}t)^2+1}\simeq{}2n-1
$
(with $n$ being the number of LL). Following such a~procedure, we have numerically reproduced the pseudodiffusive transport characteristics of a~monolayer, i.e.\ $G\simeq{}G_{\rm diff}^{(1)}$, ${\cal F}\simeq{}1/3$, and ${\cal R}\simeq{}1/15$, for any $n\geqslant{}1$. [Notice that we have set $g=0$ for clarity. When the Zeeman term is taken into account ($g=2$), the conductance approaches $G_{\rm diff}^{(1)}/2=(2/\pi)G_0W/L$ per each direction of spin, whereas the values of ${\cal F}$ and ${\cal R}$ are not altered.]

\begin{figure}[!ht]
  \centerline{
    \includegraphics[width=0.9\linewidth]{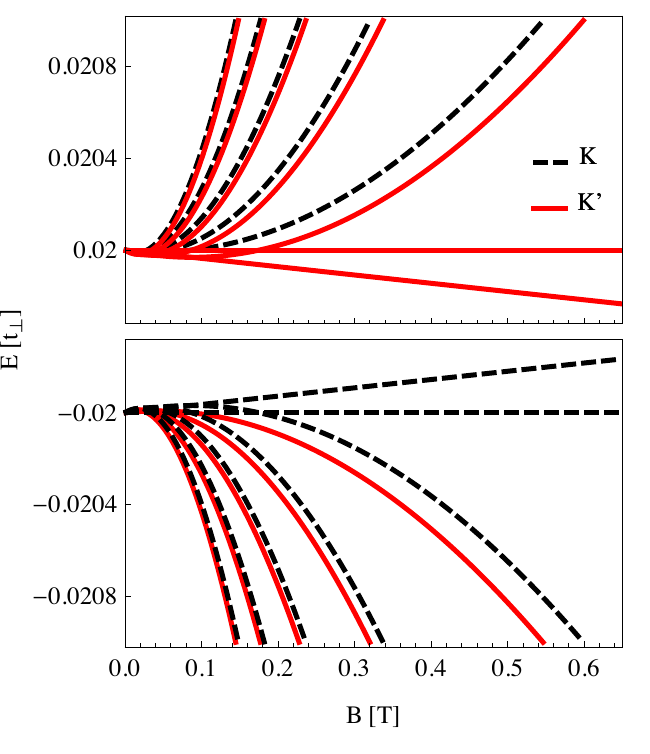}
  }
  \caption{ \label{fig:llevels}
    Magnetic field dependence of LL energies in graphene bilayer
    with a~potential bias $V=4\cdot10^{-2}t_{\perp}$ obtained from 
    Eq.\ (\ref{eq:llevels}). (The Zeeman splitting
    is not taken into account for clarity.) Notice that the states
    corresponding to different valleys ($K$ or $K'$) are exchanged
    between the conductance and the valence bands (top and bottom panels).
  }
\end{figure}

\begin{figure}[!ht]
  \centerline{
    \includegraphics[width=0.9\linewidth]{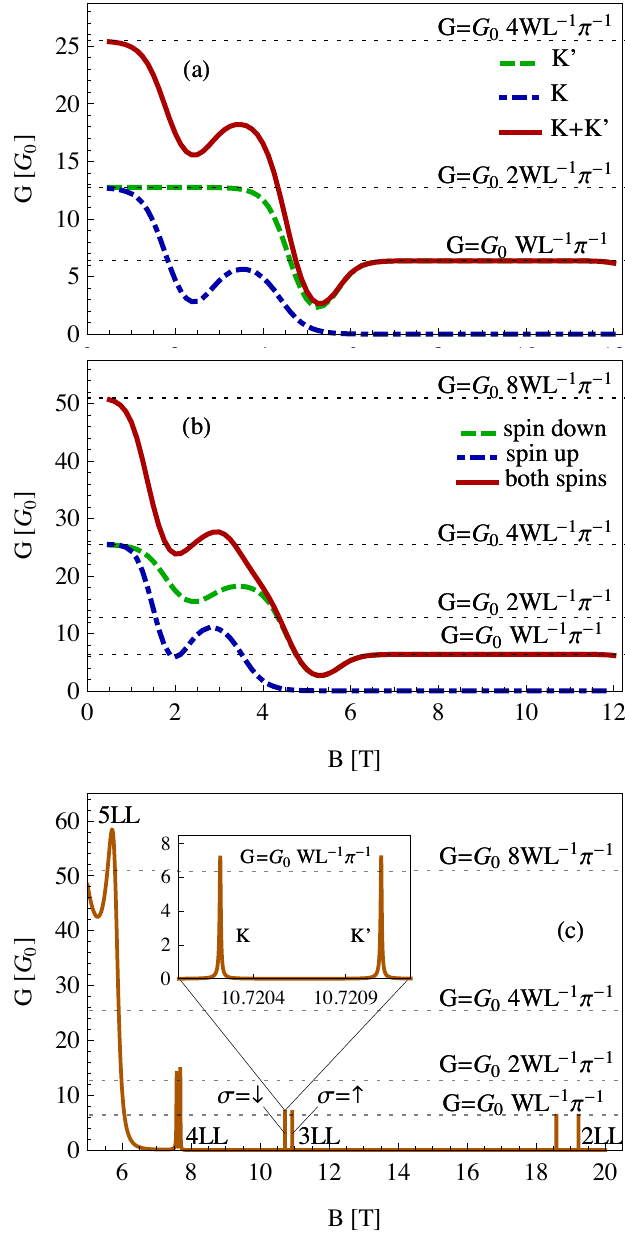}
  }
  \caption{ \label{fig:Conductance-on-0th}
    Hierarchy of Landau levels and pseudodiffusive conductance in biased graphene bilayer. (a,b) Magnetoconductance for the field-dependent doping obtained by solving Eq.\ (\ref{eq:llevels}) for $n=2$ and $m_s=-\frac{1}{2}$. Two panels show the contributions from the transmission channels corresponding to $m_s=-\frac{1}{2}$ and different valleys [panel (a)] and the conductance summed over the valleys for different directions of spin [panel (b)]. Notice the suppression of the contribution from $K$ valley and $m_s=+\frac{1}{2}$. (c) Magnetoconductance for the doping fixed at $E=0.2\,t_{\perp}$. Inset shows the separation of resonances corresponding to $K'$ and $K$ valleys for LL with $n=3$ and $m_s=-\frac{1}{2}$. 
We took $V=2\cdot{}10^{-4}\,t_{\perp}$ and $g=2$. Remaining system parameters are same as in Fig.\ \ref{fig:Transport-regimes-in}.
}
\end{figure}

\begin{figure}[!ht]
  \centerline{
    \includegraphics[width=0.9\linewidth]{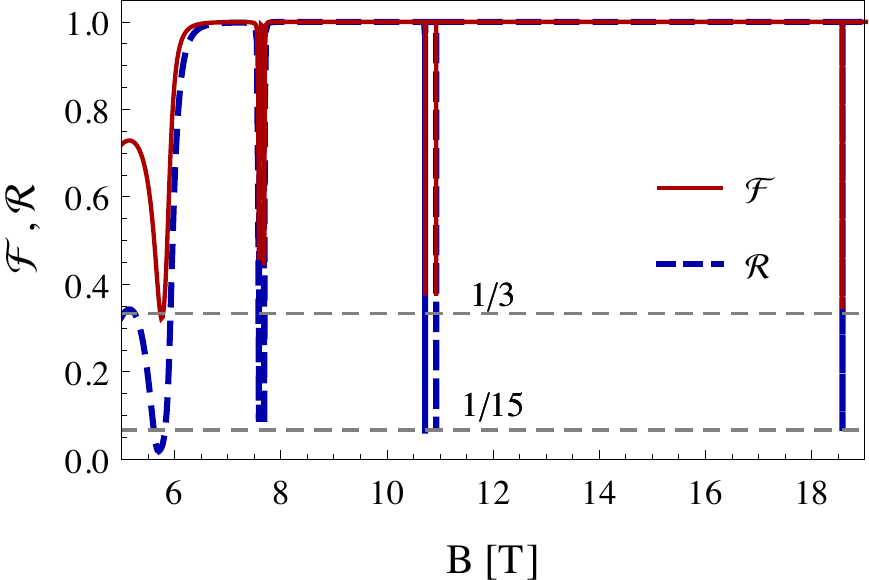}
  }
  \caption{ \label{fig:fanoir}
    Shot-noise power and the third charge-transfer cumulant, quantified by the Fano factor ${\cal F}$ (\ref{eq:ffac}) and the ${\cal R}$-factor (\ref{eq:rfac}), as functions of the magnetic field $B$. Physical parameters are same as used in Fig.\ \ref{fig:Conductance-on-0th}(c). Dashed horizontal lines mark the pseudodiffusive values ${\cal F}=1/3$ and ${\cal R}=1/15$.
  }
\end{figure}

\subsection{Graphene bilayer with nonzero bias}
We focus now on the effects appearing in the presence of a bias between the layers ($\delta\neq{}0$). Analyzing normalization conditions for the wavefunctions, one can obtain the following equation for LL energies
\begin{equation}
  \label{eq:llevels}
\varepsilon^{2}+\delta^{2}\pm\sqrt{\left(1-2\,\delta\,\varepsilon\right){}^{2}+t^{2}\,\left(\varepsilon^{2}-\delta^{2}\right)}=2\, n-1,
\end{equation}
with $n=0,1,...$. This supplements the results reported in the first paper of Ref.\ \cite{Per07}. In a~peculiar situation when $\varepsilon=\pm{}\delta$, the differential equations untangle and two additional solutions corresponding to LLs emerge, although it is not possible to find them in a~closed analytic form. Numerical values of LL energies are presented in the physical units in Fig.\ \ref{fig:llevels}.

As illustrated in Figs.\  \ref{fig:llevels} and \ref{fig:Conductance-on-0th}, the bias field lifts the valley degeneracy (see Fig.\ \ref{fig:llevels}), and thus the conductance per spin at any LL becomes two times smaller than for a~monolayer, $G\simeq{}G_{\rm diff}^{(1)}/4=(1/\pi)G_0W/L$ (see Fig.\ \ref{fig:Conductance-on-0th}). The second and third charge-transfer cumulants are still quantified by ${\cal F}\simeq{}1/3$ and ${\cal R}\simeq{}1/15$ (respectively), see Fig.\ \ref{fig:fanoir}. Also, the electron-hole symmetry is broken and the two lowest LLs ($n=0,1$) exist for electrons (or holes) only in the $K'$ (or $K$) valley, see  Fig.\ \ref{fig:llevels}.

It is worth to stress here that each LL in biased bilayer is associated with a~bunch of transmission resonances corresponding to different $k_y$-s, similarly as in the simplest case of unbiased system at the Dirac point described by Eq. (\ref{eq:trans0}). Remarkably, for the energy close to any given LL, the transmission resonances  merge in the momentum space. In fact, the wavenumber shift of $-l_{B}^{-2}L/2$ appears to provide a~reasonable approximation of the typical resonance position, regardless $\delta=0$ or $\delta\neq{}0$. For these reasons, in the numerical discussion presented in the remaining part of the paper, we suppose the mean position of transmission resonances associated with a~single LL (up to an integer multiplicity of $2\pi/W$) is given by
\begin{equation}
  \label{eq:kres}
  k_{\rm res}=\frac{2\pi}{W}\,\mbox{nint}\left(\frac{WL}{4\pi{}l_B^2}\right),
\end{equation}
where $\mbox{nint}\left(x\right)$ is the nearest integer to $x$.

\section{Effects of a~finite voltage difference or doping fluctuations \label{finvol}}
So far, we have discussed transport properties of graphene bilayer in situations when the doping $E$ is sharply-defined and the standard Landauer-B\"{u}ttiker formulas for the linear-response regime [see Eqs.\ (\ref{eq:glan}--\ref{eq:rfac})] can be applied. Such an approach may not be fully justified at high fields, when the nonzero transmission appears only at narrow doping intervals centered around LLs. For instance, the experimental results may deviate from our theoretical predictions even at zero temperature due to a~finite source-drain potential difference $V_{sd}$, as we may have (at sufficiently high $B$) $eV_{sd}\gtrsim{}{\cal W}_0$, with  ${\cal W}_0$ being the typical transmission resonance width. We also argue, that similar effects originate from slow doping fluctuations, which may occur in nanosystems when long-time measurements of the higher charge-transfer cumulants are performed. 

We now extend our analysis in order to describe the above-mentioned effects of finite $V_{sd}$ (or fluctuating doping) in a~systematic manner.  We start from presenting an empirical model describing the dependence transmission probabilities $T_{k_y}(E)$ on $k_y$ and $E$ (see Sec.\ \ref{ctrafin}). Next, theoretical predictions for ${\cal F}$ and ${\cal R}$ as functions of $V_{sd}$, arising from our model, are confronted with the corresponding results of computational experiments (see Sec.\ \ref{numer}). The evolution of statistical distribution of transmission eigenvalues $\rho(T)$ with increasing $V_{sd}$ is also briefly discussed (in Sec.\ \ref{trasta}).

\subsection{Charge-transfer cumulants at finite $V_{sd}$ \label{ctrafin}}
In the so-called shot-noise limit $eV_{sd}\gg{}k_BT$, electric charge $Q$ passing a~nanoscale graphene device during the time $\Delta{}t$ is a~random variable, a~distribution of which can be expressed via the characteristic function
\begin{equation}
  \Lambda(\chi)=\left<\exp\left(i\chi{}Q/e\right)\right>
\end{equation}
(with $\langle{}X\rangle$ denoting the expectation value of $X$), which is given by the Levitov formula \cite{Naz09}
\begin{multline}
  \ln\Lambda(\chi)=(\Delta{t}/h) \times \\
  \int_{E_0-eV_{sd}/2}^{E_0+eV_{sd}/2}dE'\,\ln\left\{\det\left[
      \boldsymbol{I}+\left(e^{i\chi}\!-\!1\right)\boldsymbol{T}(E')
    \right]\right\},
\end{multline}
where $\boldsymbol{I}$ is the identity matrix, $E_0$ is mean doping in the sample area, and we have assumed $V_{sd}>0$ for simplicity. The average charge $\langle{Q}\rangle$, as well as higher charge-transfer cumulants $\langle\langle{Q^m}\rangle\rangle\equiv\langle\,(Q-\langle{Q}\rangle)^m\,\rangle$ may be obtained by subsequent differentiation of $\ln\Lambda(\chi)$ with respect to $i\chi$ at $\chi=0$. In particular, the conductance
\begin{multline}
  \label{gfinvo}
  G(V_{sd})=\frac{\langle{Q}\rangle}{V_{sd}\Delta{t}}=\frac{e}{V_{sd}\Delta{t}}
  \left.\frac{\partial\ln\Lambda}{\partial(i\chi)}\right|_{\chi=0} \\ 
  =\frac{G_0}{eV_{sd}}
  \int_{E_0-eV_{sd}/2}^{E_0+eV_{sd}/2}dE'\,\mbox{Tr}\,\boldsymbol{T}(E') \\
  \equiv{}G_0\Big<\mbox{Tr}\,\boldsymbol{T}\Big>_{|E-E_0|\leqslant{}eV_{sd}/2},
\end{multline}
where we have identified the value of $\mbox{Tr}\,\boldsymbol{T}(E)$ averaged over the energy interval $|E-E_0|\leqslant{}eV_{sd}/2$. Eq.\ (\ref{eq:glan}) gets restored for $V_{sd}\rightarrow{}0$. Analogously,
\begin{multline}
  \label{ffinvo}
  {\cal F}(V_{sd})=\frac{\langle\langle{Q^2}\rangle\rangle}{\langle\langle{Q^2}\rangle\rangle_{\rm Poisson}} \\
\equiv\frac{\Big<\mbox{Tr}\,\left[\boldsymbol{T}\left(\boldsymbol{I}-\boldsymbol{T}\right)\right]\Big>_{|E-E_0|\leqslant{}eV_{sd}/2}}{\Big<\mbox{Tr}\,\boldsymbol{T}\Big>_{|E-E_0|\leqslant{}eV_{sd}/2}}
\end{multline}
and
\begin{multline}
  \label{rfinvo}
  {\cal R}(V_{sd})=\frac{\langle\langle{Q^3}\rangle\rangle}{\langle\langle{Q^3}\rangle\rangle_{\rm Poisson}} \\
\equiv\frac{\Big<\mbox{Tr}\,\left[\boldsymbol{T}\left(\boldsymbol{I}-\boldsymbol{T}\right)\left(\boldsymbol{I}-2\boldsymbol{T}\right)\right]\Big>_{|E-E_0|\leqslant{}eV_{sd}/2}}{\Big<\mbox{Tr}\,\boldsymbol{T}\Big>_{|E-E_0|\leqslant{}eV_{sd}/2}},
\end{multline}
where  $\langle\langle{Q^m}\rangle\rangle_{\rm Poisson}\equiv{}e^m\langle{Q}\rangle$ denotes the value of $\langle\langle{Q^m}\rangle\rangle$ for the Poissonian limit, at which all transmission probabilities $T_{k_y}(E)\ll{}1$. We notice that Eqs.\ (\ref{eq:ffac}) and (\ref{eq:rfac}) are restored for $V_{sd}\rightarrow{}0$. 

The structure of last expressions in Eqs.\ (\ref{gfinvo})--(\ref{rfinvo}) allows us to expect that the results presented in this section are also relevant for a~slightly different physical situation, namely, when $eV_{sd}\ll{\cal W}_0$, but the doping slowly fluctuates during a~measurement procedure, covering uniformly the energy interval 
\begin{equation}
  \label{deldef}
  |E-E_0|\leqslant{}{\cal W}_0\Delta/2,
\end{equation}
with $\Delta$ being the dimensionless scaling factor. For the sake of clarity, charge-transfer characteristics are hereinafter discussed as functions of $\Delta$, and the theoretical predictions for the finite-voltage situation can be immediately obtained by setting $\Delta\equiv{}eV_{sd}/{\cal W}_0$.

Our numerical results for $T_{k_y}(E)$ in case the doping $E$ is close to LL can be summarized as follows. 
\begin{enumerate}
\item[(\emph{i})]
  The transmission probability depends on the wave vector $k_y$ in a similar
manner as for a~system at zero magnetic field; i.e.\ $T_{k_y}(E)\propto\mbox{cosh}^{-2}\left[{\cal A}\left(k_{y}-k_{\rm res}\right)L\right]$, where  ${\cal A}$ is the momentum-independent empirical parameter close to unity, and $k_{\rm res}$ is given by Eq.\ (\ref{eq:kres}).
\item[(\emph{ii})]
The dependence of $T_{k_y}(E)$ on the doping $E$ can be rationalized with the Breit-Wigner distribution, characterized by ${\cal W}(k_y)$ the momentum-dependent full width at half maximum (FWHM). 

\end{enumerate}
Subsequently,
\begin{equation}
  \label{emptky}
  T_{k_y}(E)\simeq{}
  \frac{\cosh^{-2}\left[{\cal A}\left(k_{y}-k_{\rm res}\right)L\right]}{1+\left[{2(E-E_0)}/{{\cal W}(k_y)}\right]^2},
\end{equation}
where we have further assumed that the mean doping $E_0$ corresponds to the transmission maximum. Substituting the above to Eqs.\ (\ref{ffinvo},\ref{rfinvo}) and taking ${\cal W}(k_y)\simeq{}{\cal W}_0$ at~the first step, we obtain the approximating formulas for ${\cal F}$ and ${\cal R}$ in the $W\gg{}L$ limit
\begin{align}
  \overline{\cal F}\left(\Delta\right)= & 
  \frac{2}{3}-\frac{\Delta}{3(1\!+\!\Delta^{2})\arctan\Delta},
  \label{eq:FanoDel} \\
  \overline{\cal R}\left(\Delta\right)= & 
  \frac{2}{5}-\frac{\Delta}{5(1\!+\!\Delta^{2})\arctan\Delta}\left[3-\frac{4}{3\left(1\!+\!\Delta^{2}\right)}\right]. 
  \label{eq:RfacDel}
\end{align}
We observe  that $\overline{\cal F}(\Delta)$ (\ref{eq:FanoDel}) reaches its minimum at $\Delta=0$, restoring the linear-response value $\overline{\cal F}(0)=1/3$. To the contrary, the minimum of $\overline{\cal R}(\Delta)$ corresponds to a~nonzero voltage difference (or the amplitude of doping fluctuations), namely $\Delta_{\rm min}=0.34$ and ${\cal R}\left(\Delta_{\rm min}\right)=0.064$, which is slightly lower than the linear-resp-once value $\overline{\cal R}(0)=1/15$. A~striking consequence of Eqs.\ (\ref{eq:FanoDel}) and  (\ref{eq:RfacDel}) is that the second and third charge-transfer cumulants are expected to be quantum-limited also for $\Delta\rightarrow\infty$, with ${\cal F}$ and ${\cal R}$ approaching the values close to $\overline{\cal F}(\infty)=2/3$ and $\overline{\cal R}(\infty)=2/5$ (respectively), which are still significantly smaller then for the Poissonian process (${\cal F}_{\rm Poisson}={\cal R}_{\rm Poisson}=1$).

\begin{figure}[!ht]
\includegraphics[width=0.9\linewidth]{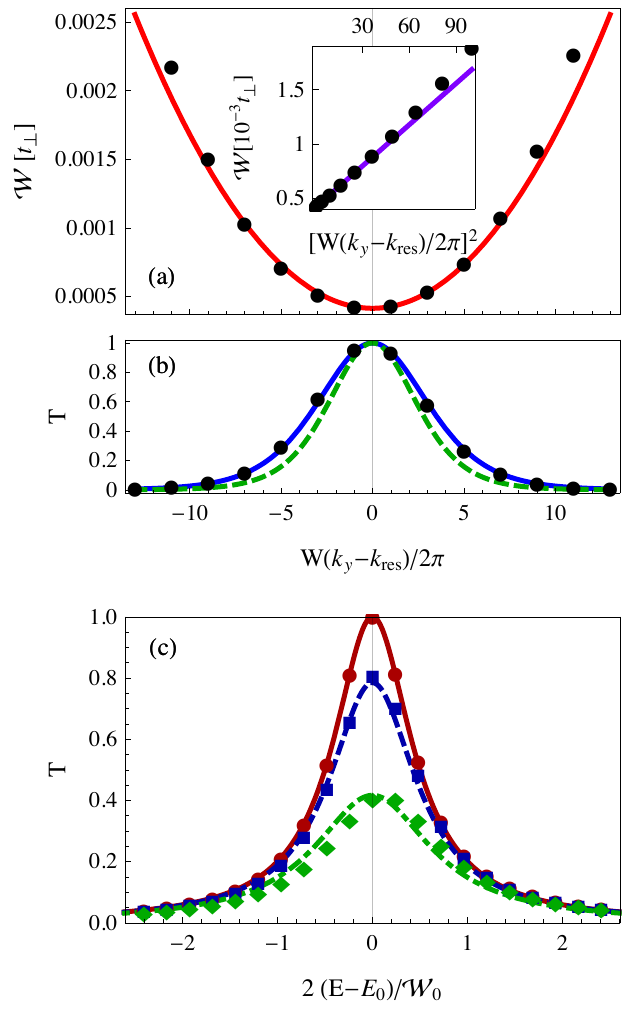}
\caption{\label{fig:Transmission-dependence-on-1}
  Transmission resonances for the second LL at $B=5\,$T and the remaining parameters same as in Fig.\ \ref{fig:Transport-regimes-in}.
(a) The resonance width ${\cal W}$ as a function of $k_y$. Datapoints are derived from the mode matching analysis, solid line depicts ${\cal W}(k_y)$ approximated by Eq.\ (\ref{wkyalp}) with the best-fitted parameters $\alpha\simeq{}1.27\cdot10^{-5}t_{\perp}$ and ${\cal W}_{0}\simeq{}4.2\cdot10^{-4}t_{\perp}$. [The inset shows the same data as a~function of $\left(k_{y}-k_{res}\right)^{2}$.] (b) Transmission probabilities for different $k_y$ and the doping fixed at $E=E_0=0.06072\,t_\perp\simeq{}0.024\,$eV. Solid (or dashed) line corresponds to Eq.\ (\ref{emptky}) with the best-fitted ${\cal A}\simeq{}0.80$ (or the fixed ${\cal A}=1$). 
(c) Transmission probability as a~function of the doping for different $k_y$: Solid, dashed, and dash-dotted line depict the values obtained from Eq.\ (\ref{emptky}) for $k_{y}\!-\!k_{res}= 0$, $4\pi/W$, and $-8\pi/W$ (with ${\cal A}\simeq{}0.80$ for all three cases).
}
\end{figure}

\begin{figure}[!ht]
\includegraphics[width=0.9\linewidth]{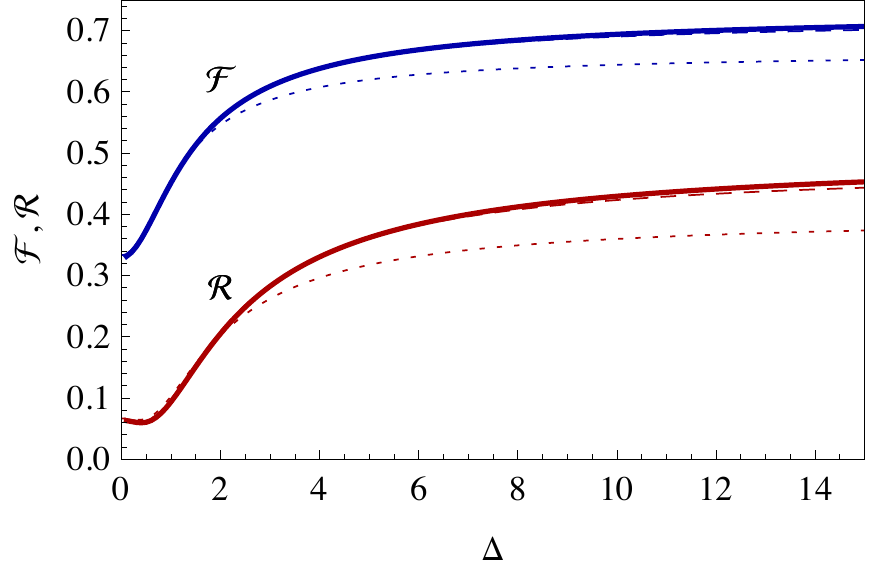}
\caption{\label{fig:FanoiR}
  ${\cal F}$ and ${\cal R}$ as functions of the voltage difference (or the amplitude of doping fluctuations), quantified by the scaling factor $\Delta$ defined via Eq.\ (\ref{deldef}). The values of $E_0$ and $B$ are same as in Fig.\ \ref{fig:Transmission-dependence-on-1}(b), the remaining parameters are same as in Fig.\ \ref{fig:Transport-regimes-in}. Solid lines depict the values obtained by calculating the averages in Eqs.\ (\ref{ffinvo},\ref{rfinvo}) numerically, for transmission matrices derived via the mode matching, whereas dashed lines corresponds to the empirical model constituted by Eqs.\ (\ref{emptky}) and (\ref{wkyalp}) with ${\cal A}=0.80$, ${\cal W}_0=4.2\cdot{}10^{-4}\,t_\perp$, and $\alpha=1.27\cdot{}10^{-5}\,t_\perp$. The approximating values of $\overline{\cal F}(\Delta)$ (\ref{eq:FanoDel}) and $\overline{\cal R}(\Delta)$ (\ref{eq:RfacDel}) are also shown (with dotted lines).
}
\end{figure}

\subsection{Numerical results \label{numer}}
Instead of employing the empirical expression for $T_{k_y}(E)$ (\ref{emptky}), one can calculate the averages in Eqs.\ (\ref{ffinvo}) and (\ref{rfinvo}) numerically, for the ensemble of actual transmission matrices $\boldsymbol{T}(E)$ obtained by repeating the mode-matching (as presented in Appendix~\ref{apptra}) for different values of $E$ sampled over a~desired energy interval \cite{compexp}. Such a~computational experiment brought us to the conclusion that $\overline{\cal F}(\Delta)$ (\ref{eq:FanoDel}) and $\overline{\cal R}(\Delta)$ (\ref{eq:RfacDel}) provide reasonable approximations of the actual ${\cal F}$ and ${\cal R}$ values for $\Delta\lesssim{}2$ only. 

Nevertheless, we find both the approximations are substantially improved when taking 
\begin{equation}
  \label{wkyalp}
  {\cal W}(k_y)\simeq{}{\cal W}_0+\alpha\left[\frac{W(k_y-k_{\rm res})}{2\pi}\right]^2,
\end{equation}
with the additional empirical parameter $\alpha$. A~comparison of ${\cal W}(k_y)$ given by Eq.\ (\ref{wkyalp}) with the values of FWHW obtained numerically is presented in Fig.\ \ref{fig:Transmission-dependence-on-1}. Next, in Fig.\ \ref{fig:FanoiR}, we compare the values of ${\cal F}$ and ${\cal R}$ obtained by means of the mode-matching analysis [solid lines], with these following from the empirical model for $T_{k_y}(E)$ constituted by Eqs.\ (\ref{emptky}) and (\ref{wkyalp}) [dashed lines]. $\overline{\cal F}(\Delta)$ (\ref{eq:FanoDel}) and $\overline{\cal R}(\Delta)$ (\ref{eq:RfacDel}) are also show in Fig.\ \ref{fig:FanoiR} [dotted lines]. Our results show that the model for $T_{k_y}(E)$ as presented, generically reproduces the actual values of ${\cal F}$ and ${\cal R}$ within $1\%$ accuracy, provided $\Delta\lesssim{}20$ and the position in the doping-field plane $(E_0,B)$ is chosen such that  ${\cal W}_0\lesssim{}10^{-3}\,t_\perp$. Morever, our prediction that the second and third charge-transfer cumulant are quantum-limited for $\Delta\rightarrow\infty$ is now further supported, and the limiting values of ${\cal F}$ and ${\cal R}$ can be approximated by 
\begin{equation}
  {\cal F}_{\infty}\simeq{} 0.7\ \ \ \ 
  \text{and} \ \ \ \ 
  {\cal R}_{\infty}\simeq{} 0.5.
\end{equation}

\begin{figure}[!ht]
  \includegraphics[width=0.9\linewidth]{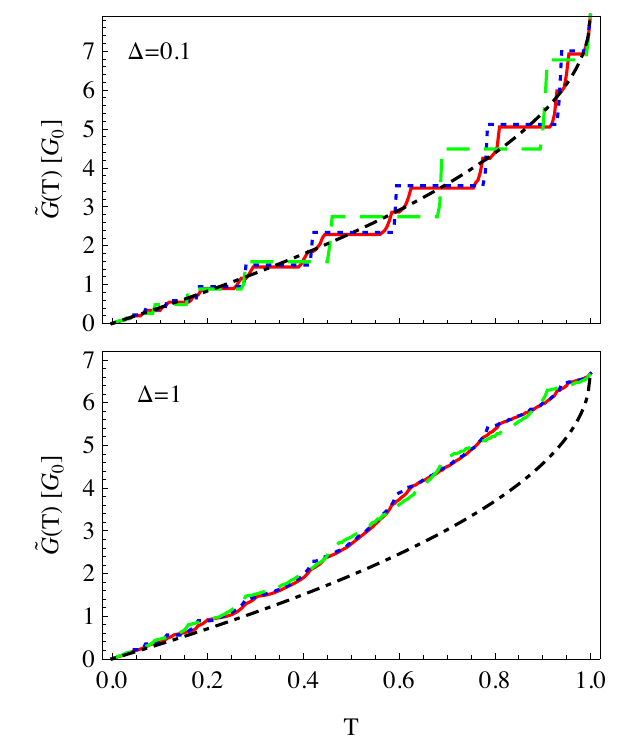}
  \caption{\label{fig:-godt}
    Partial conductance $\tilde{G}(T)$ (\ref{godtdef}) for the two values of $\Delta$ (\ref{deldef}) [specified for each panel] and the physical parameters same as in Fig.\ \ref{fig:FanoiR}. Solid lines mark the values obtained by calculating the average in Eq.\ (\ref{godtdef}) numerically for transmission matrices derived via the mode matching. Dotted lines correspond to the empirical model constituted by Eqs.\ (\ref{emptky}) and (\ref{wkyalp}) with ${\cal A}=0.80$, ${\cal W}_0=4.2\cdot{}10^{-4}\,t_\perp$, and $\alpha=1.27\cdot{}10^{-5}\,t_\perp$, whereas dashed lines present the values obtained by setting ${\cal A}=1$, $\alpha=0$, and leaving ${\cal W}_0$ same as for dotted lines. The pseudodiffusive values of $G_{\rm diff}^{(2)}(T)$ (\ref{godtdiff}) are also shown (with dot-dashed lines).
}
\end{figure}

\subsection{Transmission statistics \label{trasta}}
For the sake of completeness, we discuss now the evolution of statistical distribution of transmission eigenvalues $\rho(T)$ with the increasing voltage difference (or the amplitude of doping fluctuations), quantified by the factor $\Delta$ again [see Eq.\ (\ref{deldef})]. For the linear-response regime ($\Delta\rightarrow{}0$) such a~distribution reads \cite{Ben08,Ryc09}
\begin{equation}
  \label{rhotdiff}
  \rho_{\rm diff}^{(1,2)}(T)=\frac{2G_{\rm diff}^{(1,2)}}{\pi\sigma_0}\frac{1}{T\sqrt{1-T}},
\end{equation}
where $G_{\rm diff}^{(1,2)}$ are given by Eq.\ (\ref{gdiff12}) for graphene or its bilayer in the pseudodiffusive limit $W\ll{}L$. We further notice, that the distribution $\rho_{\rm diff}^{(1,2)}(T)$ (\ref{rhotdiff}) is normalized such that
\begin{equation}
   \int_0^1dT\,\rho_{\rm diff}^{(1,2)}(T)T = G_{\rm diff}^{(1,2)}.
\end{equation}

In our numerical discussion, the sample aspect ratio is fixed at the large but finite value $W/L=20$. (Such an approach is partly motivated by the existing experimental studies of pseudodiffusive graphene, see Refs.\ \cite{Mia07,Dic08,Dan08}.) For this reason, the total number of distinct nonzero transmission eigenvalues $T_{k_y}(E)$ in the energy interval (\ref{deldef}) is relatively small, particularly for $\Delta\lesssim{}1$. In effect, the corresponding histograms depicting $\rho(T)$ are sensitive to the choice of a~bin size. To overcome this difficulty, we introduce the so-called {\em partial conductance}
\begin{multline}
  \label{godtdef}
  \tilde{G}(T)/G_0=\int_0^TdT'\rho(T')T' \\
  \equiv\int_0^TdT'\Big<\mbox{Tr}\,\left[
    \boldsymbol{T}\,\delta_{\xi\rightarrow{}0}\left(\boldsymbol{T}\!-\!T'\boldsymbol{I}\right)
  \right]\Big>_{|E-E_0|\leqslant{}{\cal W}_0\Delta/2},
\end{multline}
where $\delta_{\xi\rightarrow{}0}(\boldsymbol{M})$ is an analytic representation of the Dirac delta function with a~matrix argument $\boldsymbol{M}$. [For instance, $\tilde{G}(1)$ reproduces the conductance as given by Eq.\ (\ref{gfinvo}).] In the pseudodiffusive limit, we have
\begin{multline}
  \label{godtdiff}
  \tilde{G}_{\rm diff}^{(1,2)}(T)=G_0\int_0^TdT'\rho_{\rm diff}^{(1,2)}(T')T'\\
  =G_{\rm diff}^{(1,2)}\left(1-\sqrt{1-T}\right).
\end{multline}

In Fig.\ \ref{fig:-godt}, we compare $\tilde{G}(T)$ obtained from Eq.\ (\ref{godtdef}) utilizing three different numerical approaches, in analogy to the earlier presentation of Fig.\ \ref{fig:FanoiR}. First, the average in Eq.\ (\ref{godtdef}) is  calculated for actual transmission matrices derived via the mode-matching [solid lines]. Next, the empirical model constituted by Eqs.\ (\ref{emptky}) and (\ref{wkyalp}) [dotted lines] and its simplified version obtained by setting ${\cal W}(k_y)\simeq{}{\cal W}_0$ [dashed lines] are employed. The values of $G_{\rm diff}^{(2)}(T)$ (\ref{godtdiff}) are also shown in Fig.\ \ref{fig:-godt} [dot-dashed lines]. Our results show that the actual distribution of transmission eigenvalues $\rho(T)$ may follow the pseudodiffusive distribution $\rho_{\rm diff}^{(2)}(T)$ (\ref{rhotdiff}) only if the doping energy is adjusted rather closely to LL ($\Delta=0.1$). When doping fluctuations get larger ($\Delta=1$), a~significant deviation of  $\rho(T)$ from $\rho_{\rm diff}^{(2)}(T)$ is observed, due to the enhanced contribution of low transmission eigenvalues. In both cases, the agreement with the empirical model presented earlier [see Eqs.\ (\ref{emptky}) and (\ref{wkyalp})] is excellent.

\section{Influence of indirect interlayer hopping integrals}

\begin{figure}[!ht]
\centerline{
  \includegraphics[width=0.9\linewidth]{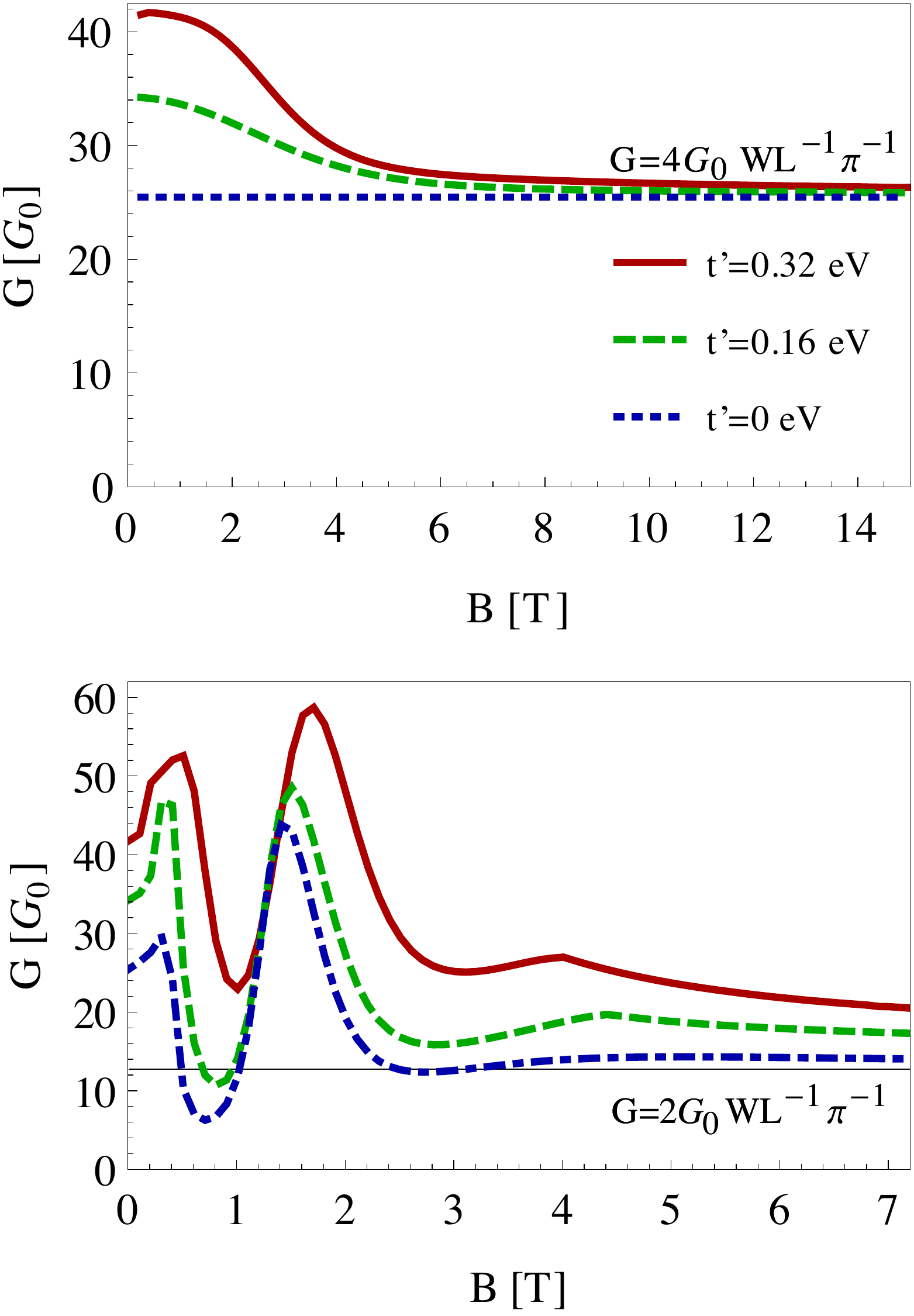}
}
\caption{\label{fig:numericalg}
  Magnetoconductance of unbiased graphene bilayer (per one direction of spin) for different values of the next-nearest neighbor interlayer hopping $t'$. The field-dependent doping is adjusted to follow the transmission maxima for $n=0$ (top panel) and $n=2$ (bottom panel) Landau levels. The system parameters are same as used in Fig.\ \ref{fig:Transport-regimes-in}.}
\end{figure}

\begin{figure}[!ht]
\centerline{
  \includegraphics[width=0.9\linewidth]{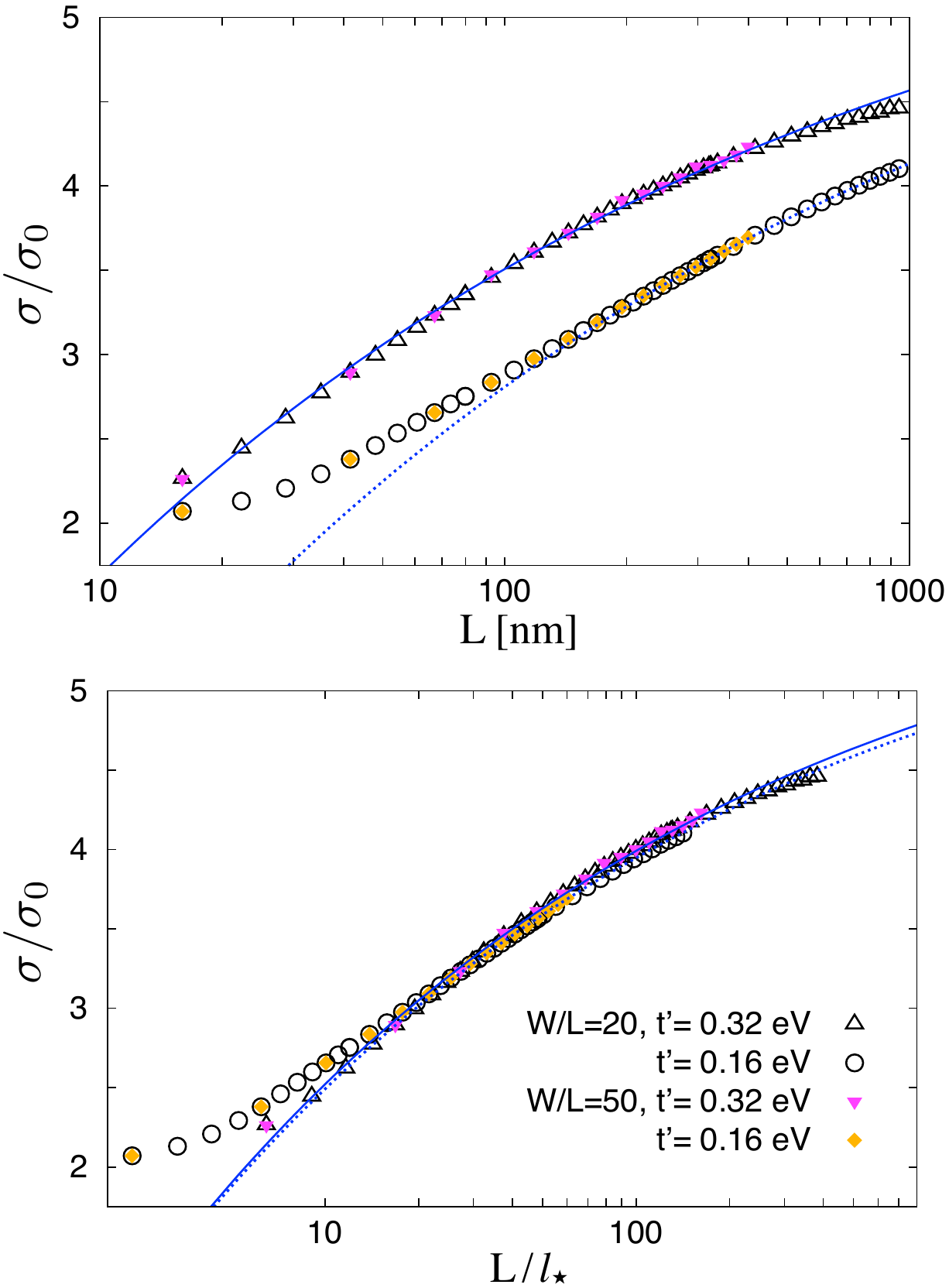}
}
\caption{\label{fig:gscaling}
  Minimal conductivity of unbiased bilayer as a~function of the sample length $L$ for $W/L=20$ (open symbols) and $W/L=50$ (solid symbols). The value of $t'$ is specified for each dataset. Lines show the best fitted power-law relations (\ref{powerlaws}) with parameters given by Eq.\ (\ref{param32}) [solid lines] and Eq.\ (\ref{param16}) [dashed lines]. Top panel shows the raw data. In the bottom panel the data sets are shifted horizontally to demonstrate the universal behavior for $L/l_\star\gg{}1$. 
}
\end{figure}

Theoretical calculations based on the Kubo formula \cite{Cse07} show that the minimal conductivity of ballistic graphene bilayer may be unstable with respect to indirect interlayer hopping integrals \cite{Kuz09}, which are neglected in the Hamiltonian (\ref{eq:hamiltonian1}). At zero field and zero bias situation ($B=V=0$), the minimal conductivity is predicted to be $(24/\pi)\,e^2/h=6\sigma_0$ (i.e, {\em 6~times} larger than the conductivity of a~monolayer) for arbitrarily small indirect interlayer hoppings. In the absence of such hoppings, the Kubo conductivity drops back to $2\sigma_0$, what is attributed to the disappearance of additional Fermi surface pockets at low energies \cite{Mac13,Wan12}. (We notice here, that the effect has no high-frequency analog, what beautifully manifests itself by direct scaling of visible light absorption with the number of layers, see Ref.\ \cite{Nai08}.) The experimental value of $\sigma_{xx}\simeq{}5\sigma_0$ \cite{May11} is close to the prediction of Ref.\ \cite{Cse07}, with a~small deviation which may be related to several factors, such as a~finite system size, the presence of disorder, electron-phonon coupling, or electron-electron interactions, not taken into account by existing theory in a~rigorous manner. Additionally, the values following from the Kubo formula are known to be sensitive to the order in which certain limits are taken \cite{Zie07}. For these reasons, an independent calculation employing the Landauer-B\"{u}ttiker formalism for a~ballistic system of a~given length $L$ and a~width $W$, allowing one at least to identify the possible effects of a~finite system size, is desired. 

The Hamiltonian for $K$ valley (\ref{eq:hamiltonian1}) is now replaced by
\begin{equation}
  \label{eq:hamiltonian1p}
  H'=\left(\begin{array}{cccc}
      U_{1}(x) & \pi_{x}\!+\!i\pi_{y} & t_{\bot} & 0\\
      \pi_{x}\!-\!i\pi_{y} & U_{1}(x) & 0 & \pi_{x}'\!+\!i\pi_{y}'\\
      t_{\bot} & 0 & U_{2}(x) & \pi_{x}\!-\!i\pi_{y}\\
      0 &  \pi_{x}'\!-\!i\pi_{y}' & \pi_{x}\!+\!i\pi_{y} & U_{2}(x)
    \end{array}\right),
\end{equation}
where $\pi_j'=(t'/t_0)\,\pi_j$ with $j=1,2$, $t_0=\frac{2}{3}\sqrt{3}\hbar{}v_F/a$ is the nearest neighbor hopping in a~single layer defined via the Fermi velocity and the lattice spacing $a=0.246\,$nm, $t'$ is the next-nearest neighbor interlayer hopping \cite{hoppingfoo}, and the remaining symbols are same as in Eq.\ (\ref{eq:hamiltonian1}). Next, the Dirac equation $H'\psi=E\psi$ is solved numerically for the sample area $0<x<L$, separately for each value of the transverse wavenumber $k_y=2\pi{}n/W$ (with $n=0,\pm{}1,\pm{}2,\dots$) following from the periodic boundary conditions. The mode matching analysis is then carried out as reported in Appendix~\ref{apptra}. Although the wavefunctions for $t'\neq{}0$ can still be obtained analytically in some particular situations (and will be given elsewhere), the compact-form expressions for transmission eigenvalues $T_{k_y}(E)$, such as given by Eq.\ (\ref{eq:trans0}), are now unavailable even for the simplest $E=0$ and $B=0$ case.

The numerical results are presented in Figs.\ \ref{fig:numericalg} and \ref{fig:gscaling}, where we have further limited our discussion to the case of a~zero bias between the layers ($V=0$) and to the limit of wide samples ($W\gg{}L$).

In Fig.\ \ref{fig:numericalg}, we demonstrate (as a~proof of principle) that indirect interlayer hoppings play no role at high magnetic fields, for which $l_B\ll{}L$, and the transmission resonances via individual LLs are well-defined. In such a~limit, the conductance per one direction of spin approaches the value of $G_{\rm diff}^{(1)}=(4/\pi)\,G_0W/L$ for the Dirac point (see top panel) or $G_{\rm diff}^{(1)}/2$ for higher LLs (see bottom panel for $n=2$ case) in consistency with the results reported in Sec.\ \ref{traunbigra} for the $t'=0$ case. In the opposite limit of $B\rightarrow{}0$, the zero-energy conductance is enhanced by a~factor of $1.6$ for $t'=0.32\,$eV [or $1.3$ for $t'=0.16\,$eV] above the pseudodiffusive value, what is significantly smaller than a~$t'$-independent factor $3$ predicted by Ref.\ \cite{Cse07}. The pseudodiffusive values of ${\cal F}$ and ${\cal R}$ are unaffected by $t'$. (We further notice that the numerical results presented in Fig.\ \ref{fig:numericalg} correspond to $W/L=20$ and $L=48\,\hbar{}v_F/t_\perp\simeq{}77\,$nm.)

To further understand the nature of this clear discrepancy between the results obtained employing the Kubo formula and the Landauer-B\"{u}ttiker formalism, we analyze numerically the ballistic conductivity at $E=B=0$ as a~function of $L$ (see Fig.\ \ref{fig:gscaling}). We find that the conductivity is no longer universal for $t'>0$, but slowly grows with $L$, and can be approximated (for large $L$) within a~power law relation
\begin{equation}
  \label{powerlaws}
  \sigma(L)=\sigma_\infty\left[\, 
    1 - \left(\frac{l_\star}{L}\right)^\gamma 
  \,\right].
\end{equation} 
Least-square fitted parameters in Eq.\ (\ref{powerlaws}) are 
\begin{multline}
  \label{param32}
  \sigma_\infty\!\!=\!6.05\sigma_0,\ l_\star\!\!=\!2.5\,\text{nm},\ 
  \gamma\!=\!0.23 \ \ 
  \text{for}\ t'\!=\!0.32\,\text{eV},
\end{multline}
and 
\begin{multline}
  \label{param16}
  \sigma_\infty\!\!=\!6.0\,\sigma_0,\ l_\star\!\!=\!6.7\,\text{nm},\ 
  \gamma\!=\!0.23 \ \ 
  \text{for }\ t'\!=\!0.16\,\text{eV},
\end{multline}
with the standard deviations not exceeding $1\%$ in all cases. We observe that only the parameter $l_\star$ significangly varies with $t'$. Replotting the conductivity as a~function of the dimensionless variable $L/l_\star$ (see bottom panel in Fig.\ \ref{fig:gscaling}) shows the universal nature of the length-dependence of the conductivity. 

Although the Landauer-B\"{u}ttiker conductivity approaches the value of $\sigma_\infty\simeq{}6\sigma_0$ for $L\rightarrow\infty$, restoring the results of Ref.\ \cite{Cse07}, the values of $\sigma(L)$ following from Eq.\ (\ref{powerlaws}) for typical lengths of ballistic samples used in the experiments are still significantly smaller that $6\sigma_0$. In particular, using the parameters given by Eqs.\ (\ref{param32},\ref{param16}) for an extrapolation, one gets $\sigma(L\!=\!1\,\mu\text{m})=4.1-4.6\,\sigma_0$ and  $\sigma(L\!=\!10\,\mu\text{m})=4.9-5.2\,\sigma_0$, where the upper (lower) limit corresponds to $t'=0.32\,$eV ($t'=0.16\,$eV). Therefore, the fact that experimental values of the minimal conductivity \cite{May11} are noticeably smaller than the prediction of Ref.\ \cite{Cse07} may be predominantly caused by finite system sizes, with only a~secondary role played by the disorder or many-body effects.

\section{Conclusions}
We have calculated the conductance $G$, the Fano factor ${\cal F}$, and the factor ${\cal R}$ quantifying the third charge-transfer cumulant, for a~ballistic strip in graphene bilayer, in the presence of bias between the layers and strong magnetic fields. Our results show that the so-called pseudodiffusive charge-transport regime appears generically for a~sample with large aspect ratio ($W\gg{}L$) not only at the Dirac point (DP), but also in the vicinity of any Landau level (LL). However, the conductivity $\sigma=GL/W$ in the pseudodiffusive regime is not always equal to $2\sigma_0$ (with $\sigma_0=(4/\pi)\,e^2/h$ being the conductivity of a~monolayer) as predicted for a~zero-field and zero-bias situation by Snyman and Beenakker \cite{Sny07}, but takes quantized values of $s\sigma_0/2$, with the prefactor $s=1$, $2$, $4$ or $8$, depending whether each of spin, valley, and layer degeneracies is present or absent (see Table~\ref{tab:pseudodiffusive}). 

Other charge-transfer characteristics studied are insensitive to the splittings of degeneracies, leading to ${\cal F}\simeq{}1/3$ and ${\cal R}\simeq{}1/15$ in any case the pseudodiffusive regime is approached. This observation is further supported with statistical analysis of the distribution of transmission eigenvalues, which follows the corresponding distribution for a~diffusive wire, provided the sample doping is kept in a~vicinity of DP or LL.

Additionally, the analysis is extended beyond the standard linear-response regime; i.e., we considered the effects of a~finite voltage difference or slow doping fluctuations. Numerical analysis of transmission matrices obtained via the mode-matching for the Dirac equation at different dopings allows us to propose an empirical model for transmission probabilities, which is then used to rationalize the dependence of charge-transfer characteristics on the voltage difference (or the amplitude of doping fluctuations). Probably, the most remarkable feature of these results is that both the shot-noise power and the third charge-transfer cumulant are predicted to be quantum-limited also for large doping fluctuations, leading to ${\cal F}$ and ${\cal R}$ approaching the limiting values of ${\cal F}_\infty\simeq{}0.5$ and ${\cal R}_\infty\simeq{}0.7$. 

Finally, we have discussed the influence of indirect interlayer hoppings (quantified by $t'$) on the conductance and other charge-transfer characteristics. The results show that such hoppings may only affect the conductance at zero or weak magnetic fields. At stronger fields, when LLs are formed, the behavior earlier identified for $t'=0$ is restored. Surprisingly, for $t'\neq{}0$ the zero-field zero-bias conductivity at the Dirac point is neither equal to $2\sigma_0$ \cite{Sny07} nor $6\sigma_0$ \cite{Cse07}, but grows monotonically with the system length, taking the values from an interval $2\sigma_0<\sigma(L)<6\sigma_0$. A very slow convergence to the upper conductivity limit is observed for large $L$ and can be rationalized as $\sigma(L)\simeq{}6\sigma_0\left[1-(l_\star/L)^{-\gamma}\right]$, with a~$t'$-independent exponent $\gamma\simeq{}0.23$. The characteristic length $l_\star$ is of the order of nanometers and strongly depends on $t'$, offering a~possibility to determine the effective value of $t'$ solely by the minimal conductivity measurement at fixed $L\ll{}W$. 

\begin{table}[!ht]
\caption{\label{tab:pseudodiffusive}
  The degeneracy prefactors occurring in the expression for pseudodiffusive conductance $G=se^2/(\pi{}h)\times{}W/L$ for graphene or its bilayer in different physical situations. Indexes $\sigma$, $v$ and $l$ marks the degeneracies associated with spin, valley, and layer degrees of freedom (respectively). 
}
\begin{tabular}{c|c|c|c}
\hline\hline
Degeneracy, & $B=0$ & \multicolumn{2}{c}{$B\neq{}0$}  \\
$s$ & & $0$-th LL & Other LLs \\ 
\hline 
Monolayer & $4_{(\sigma,v)}$ & $2_{(v)}$ & $2_{(v)}$ \\
\hline 
Bilayer, $V=0$ & $8_{(\sigma,v,l)}$\footnote{ 
  This particular value applies in the absence of indirect interlayer hopping ($t'=0$) only.
} & $4_{(v,l)}$ & $2_{(v)}$ \\
\hline 
Bilayer, $V\neq0$ & $4_{(\sigma,v)}$ & $1$ & $1$ \\
\hline\hline
\end{tabular} 
\end{table}

\section*{Acknowledgements}
We thank to Patrik Recher for helpful comments and the correspondence.
The work was supported by the National Science Centre of Poland (NCN) via Grant No.\ N--N202--031440, and partly by Foundation for Polish Science (FNP) under the program TEAM {\em ``Correlations and coherence in quantum materials and structures (CCQM)''}. G.R.\ acknowledges the support from WIKING project. Computations were partly performed using the PL-Grid infrastructure.


\appendix

\begin{widetext}

\section{Wavefunctions \label{appfun}}

In this Appendix we present the wavefunctions of a charge carrier in a~carbon bilayer at the Dirac point as well as at finite dopings, in the presence of a~uniform magnetic field.

\subsection{The Dirac point  ($\varepsilon=\delta=0$)}
A general solution of Eq.\ (\ref{eqfiafib}) for $\varepsilon=\delta=0$ has the form of a~linear combination of four independent spinors with arbitrary coefficients $C_1,\dots,C_4$, namely
\begin{eqnarray}
\left(\begin{array}{c}
\phi_{A_{1}}(x)\\
\phi_{B_{1}}(x)\\
\phi_{B_{2}}(x)\\
\phi_{A_{2}}(x)
\end{array}\right) & = & 
C_1\left(\begin{array}{c} 
    f_{B,k_y}(x) \\ 0 \\ 0 \\ -t_{\perp}xf_{B,k_y}(x)
  \end{array}\right) +
C_2\left(\begin{array}{c} 
    0 \\ \bar{f}_{B,k_y}(x) \\ 0 \\ 0 
  \end{array}\right) +
C_3\left(\begin{array}{c} 
    0 \\ -t_{\perp}x\bar{f}_{B,k_y}(x) \\ \bar{f}_{B,k_y}(x) \\ 0 
  \end{array}\right) +
C_4\left(\begin{array}{c} 
    0 \\ 0 \\ 0 \\ f_{B,k_y}(x)
  \end{array}\right),
\label{eq:ffaloweBdomzerowe}
\end{eqnarray}
where $f_{B,k_y}(x)=\exp(l_{B}^{-2}\, x^{2}/2-x\, k_{y})$ and $\bar{f}_{B,k_y}(x)=1/f_{B,k_y}(x)$.

\subsection{Finite dopings ($\varepsilon\neq{}0$ or $\delta\neq{}0$)}
In the case of finite dopings ($\varepsilon\neq{}0$ or $\delta\neq{}0$) we have two pairs of solutions, hereinafter labeled as $\phi_{1,\alpha}^{\pm}$ and $\phi_{2,\alpha}^{\pm}$ (with the signs $\pm$ related to the two subbands), which are given by
\begin{equation}
\begin{array}{ccc}
\phi_{1,A_{1}}^{\pm}\,(\varepsilon,\delta;\,\xi) & = & \mbox{e}^{-\xi^{2}/4}{}_{1}\mbox{F}_{1}\left(\frac{1-2\zeta_{\pm}}{4};\,\frac{1}{2};\,\frac{\xi^{2}}{2}\right)\\
\phi_{2,A_{1}}^{\pm}\,(\varepsilon,\delta;\,\xi) & = & \xi\,\mbox{e}^{-\xi^{2}/4}{}_{1}\mbox{F}_{1}\left(\frac{3-2\zeta_{\pm}}{4};\,\frac{3}{2};\,\frac{\xi^{2}}{2}\right),\\
\phi_{1,B_{1}}^{\pm}\,(\varepsilon,\delta;\,\xi) & = & \left(1+2\,\zeta_{\pm}\right)\left[\left(\delta+\varepsilon\right)\sqrt{2}\right]^{-1}\,\xi\,\mbox{e}^{-\xi^{2}/4}{}_{1}\mbox{F}_{1}\left(\frac{1-2\zeta_{\pm}}{4};\,\frac{3}{2};\,\frac{\xi^{2}}{2}\right),\\
\phi_{2,B_{1}}^{\pm}\,(\varepsilon,\delta;\,\xi) & = & \left[\left(\delta+\varepsilon\right)\,3\sqrt{2}\right]^{-1}\,\mbox{e}^{-\xi^{2}/4}\left\{ \left(3+2\,\zeta_{\pm}\right)\,\xi^{2}\,{}_{1}\mbox{F}_{1}\left(\frac{3-2\zeta_{\pm}}{4};\,\frac{5}{2};\,\frac{\xi^{2}}{2}\right)\right.\\
 &  & \left.-6\,{}_{1}\mbox{F}_{1}\left(\frac{3-2\zeta_{\pm}}{4};\,\frac{3}{2};\,\frac{\xi^{2}}{2}\right)\right\} ,\\
\phi_{1,B_{2}}^{\pm}\,(\varepsilon,\delta;\,\xi) & = & \left(\delta+\varepsilon\right)^{-1}\,\alpha_{\pm}\,\mbox{e}^{-\xi^{2}/4}{}_{1}\mbox{F}_{1}\left(\frac{1-2\zeta_{\pm}}{4};\,\frac{1}{2};\,\frac{\xi^{2}}{2}\right),\\
\phi_{2,B_{2}}^{\pm}\,(\varepsilon,\delta;\,\xi) & = & \left(\delta+\varepsilon\right)^{-1}\,\alpha_{\pm}\,\xi\,\mbox{e}^{-\xi^{2}/4}{}_{1}\mbox{F}_{1}\left(\frac{3-2\zeta_{\pm}}{4};\,\frac{3}{2};\,\frac{\xi^{2}}{2}\right),\\
\phi_{1,A_{2}}^{\pm}\,(\varepsilon,\delta;\,\xi) & = & \left[\left(1-2\,\zeta_{\pm}\right)\,\alpha_{\pm}/\sqrt{2}\right]\,\xi\,\mbox{e}^{-\xi^{2}/4}{}_{1}\mbox{F}_{1}\left(\frac{5-2\zeta_{\pm}}{4};\,\frac{3}{2};\,\frac{\xi^{2}}{2}\right),\\
\phi_{2,A_{2}}^{\pm}\,(\varepsilon,\delta;\,\xi) & = & \alpha_{\pm}\left[\left(\delta^{2}-\varepsilon^{2}\right)3\sqrt{2}\right]^{-1}\,\mbox{e}^{-\xi^{2}/4}\,\left\{ 6\,\left(1+\xi^{2}\right)\,{}_{1}\mbox{F}_{1}\left(\frac{3-2\zeta_{\pm}}{4};\,\frac{3}{2};\,\frac{\xi^{2}}{2}\right)\right.\\
 &  & \left.-\xi^{2}\,\left(3+2\,\zeta_{\pm}\right)\,{}_{1}\mbox{F}_{1}\left(\frac{3-2\zeta_{\pm}}{4};\,\frac{5}{2};\,\frac{\xi^{2}}{2}\right)\right\} ,
\end{array}\label{eq:ffalowebialyer}
\end{equation}
where $\xi=\sqrt{2}\left(l_{B}^{-1}x+l_{B}\, k_{y}\right)$,
$\alpha_{\pm}=\left[\left(\delta+\varepsilon\right)^{2}-1-2\,\zeta_{\pm}\right]/t$, $\zeta_{\pm}=\frac{1}{2}\left[\varepsilon^{2}+\delta^{2}\pm\sqrt{(1-2\delta\varepsilon)^{2}+t^{2}\left(\varepsilon^{2}-\delta^{2}\right)}\right]$, $_pF_q(a_1,\dots,a_p;b_1,\dots,b_q;z)$ denotes the generalized hypergeometric function \cite{Olv10}, and the remaining symbols are same as in Eq.\ (\ref{eqfiafib}) in the main text.

\section{Transmission eigenvalues \label{apptra}}
Using wavefunctions of the form $\psi=\left(\phi_{A_{1}},i\phi_{B_{1}},\phi_{B_{2}},i\phi_{A_{2}}\right)^{T}$, one can write the charge-conservation conditions for a~strip of width $W$ and length $L$ (see Fig.\ \ref{setupfig}) in graphene bilayer as follows
\begin{equation}
\begin{array}{ccc}
\psi_{R,{\rm I}}^{\pm}(x_{0})+r_{p}^{\pm}\psi_{L,{\rm I}}^{+}(x_{0})+r_{n}^{\pm}\,\psi_{L,{\rm I}}^{-}(x_{0}) & = & \psi_{\rm II}(x_{0}),\\
t_{p}^{\pm}\,\psi_{R,{\rm III}}^{+}(x_{1})+t_{n}^{\pm}\,\psi_{R,{\rm III}}^{-}(x_{1}) & = & \psi_{\rm II}(x_{1}),
\end{array}
\end{equation}
where we set $x_{0}=0$, $x_{1}=L$. The lower indexes $R$ and $L$ refer to the solutions moving to the right or left (respectively), whereas the indexes I, II and III refer to left contact, sample and right contact. The upper indexes $\pm$ refer to the two subbands, and $r_p^{\pm}$, $r_l^{\pm}$ ($t_p^{\pm}$, $t_l^{\pm}$) denote the corresponding reflection (transmission) amplitudes. We further suppose that the functions $\psi_R^{\pm}$, $\psi_L^{\pm}$ in regions I and III are normalized to carry a unit current. 

Taking the limit of $|U_{\infty}|\rightarrow\infty$ [i.e., choosing the functions  $\psi_R^{\pm}$, $\psi_L^{\pm}$ for regions I and III as given by Eqs.\ (\ref{psillim}) and (\ref{psirlim}) in the main text] we obtain the following system of linear equations

\begin{equation}
\begin{array}{c}
\left(\begin{array}{cccccccc}
1 & -1 & \phi_{1,A_{1}}^{+}\,(\varepsilon,\delta;\,\xi_{0}) & \phi_{1,A_{1}}^{-}\,(\varepsilon,\delta;\,\xi_{0}) & \phi_{2,A_{1}}^{+}\,(\varepsilon,\delta;\,\xi_{0}) & \phi_{2,A_{1}}^{-}\,(\varepsilon,\delta;\,\xi_{0}) & 0 & 0\\
-1 & 1 & i\phi_{1,B_{1}}^{+}\,(\varepsilon,\delta;\,\xi_{0}) & i\phi_{1,B_{1}}^{-}\,(\varepsilon,\delta;\,\xi_{0}) & i\phi_{2,B_{1}}^{+}\,(\varepsilon,\delta;\,\xi_{0}) & i\phi_{2,B_{1}}^{-}\,(\varepsilon,\delta;\,\xi_{0}) & 0 & 0\\
-1 & -1 & \phi_{1,B_{2}}^{+}\,(\varepsilon,\delta;\,\xi_{0}) & \phi_{1,B_{2}}^{-}\,(\varepsilon,\delta;\,\xi_{0}) & \phi_{2,B_{2}}^{+}\,(\varepsilon,\delta;\,\xi_{0}) & \phi_{2,B_{2}}^{-}\,(\varepsilon,\delta;\,\xi_{0}) & 0 & 0\\
1 & 1 & i\phi_{1,A_{2}}^{+}\,(\varepsilon,\delta;\,\xi_{0}) & i\phi_{1,A_{2}}^{-}\,(\varepsilon,\delta;\,\xi_{0}) & i\phi_{2,A_{2}}^{+}\,(\varepsilon,\delta;\,\xi_{0}) & i\phi_{2,A_{2}}^{-}\,(\varepsilon,\delta;\,\xi_{0}) & 0 & 0\\
0 & 0 & \phi_{1,A_{1}}^{+}\,(\varepsilon,\delta;\,\xi_{1}) & \phi_{1,A_{1}}^{-}\,(\varepsilon,\delta;\,\xi_{1}) & \phi_{2,A_{1}}^{+}\,(\varepsilon,\delta;\,\xi_{1}) & \phi_{2,A_{1}}^{-}\,(\varepsilon,\delta;\,\xi_{1}) & 1 & -1\\
0 & 0 & i\phi_{1,B_{1}}^{+}\,(\varepsilon,\delta;\,\xi_{1}) & i\phi_{1,B_{1}}^{-}\,(\varepsilon,\delta;\,\xi_{1}) & i\phi_{2,B_{1}}^{+}\,(\varepsilon,\delta;\,\xi_{1}) & i\phi_{2,B_{1}}^{-}\,(\varepsilon,\delta;\,\xi_{1}) & 1 & -1\\
0 & 0 & \phi_{1,B_{2}}^{+}\,(\varepsilon,\delta;\,\xi_{1}) & \phi_{1,B_{2}}^{-}\,(\varepsilon,\delta;\,\xi_{1}) & \phi_{2,B_{2}}^{+}\,(\varepsilon,\delta;\,\xi_{1}) & \phi_{2,B_{2}}^{-}\,(\varepsilon,\delta;\,\xi_{1}) & -1 & -1\\
0 & 0 & i\phi_{1,A_{2}}^{+}\,(\varepsilon,\delta;\,\xi_{1}) & i\phi_{1,A_{2}}^{-}\,(\varepsilon,\delta;\,\xi_{1}) & i\phi_{2,A_{2}}^{+}\,(\varepsilon,\delta;\,\xi_{1}) & i\phi_{2,A_{2}}^{-}\,(\varepsilon,\delta;\,\xi_{1}) & -1 & -1
\end{array}\right)\cdot\left(\begin{array}{c}
r_{p}^{\pm}\\
r_{n}^{\pm}\\
C_{1}^{\pm}\\
C_{2}^{\pm}\\
C_{3}^{\pm}\\
C_{4}^{\pm}\\
t_{p}^{\pm}\\
t_{n}^{\pm}
\end{array}\right)=\left(\begin{array}{c}
\mp1\\
\mp1\\
1\\
1\\
0\\
0\\
0\\
0
\end{array}\right)\end{array},
\end{equation}
where $\xi_{0}=\sqrt{2}\,l_{B}\,k_{y}$, $\xi_1=\sqrt{2}\left(l_{B}^{-1}L+l_{B}\, k_{y}\right)$, and the remaining symbols are same as used in Appendix \ref{appfun}.
In turn, the transmission matrix for the $K$ valley and the transverse momentum fixed at $k_y$ is of the form:

\begin{equation}
\label{t1block}
  \mathbf{t}_{K,k_y}\left(\varepsilon,\delta\right)=\left(\begin{array}{cc}
      t_{p}^{+} & t_{n}^{+}\\
      t_{p}^{-} & t_{n}^{-}
    \end{array}\right).
\end{equation}
[Notice that the dependence on the $z$-component of spin $m_s$ is incorporated in $\varepsilon$, see Eq.\ (\ref{eqfiafib}) in the main text.] The transmission matrix for the $K'$ valley can be obtained from an~analogous procedure, starting from the wavefunction $\psi'=\left(\phi_{A_{1}},-i\phi_{B_{1}},\phi_{B_{2}},-i\phi_{A_{2}}\right)^{T}$, with the components given by Eq.\ (\ref{eq:ffalowebialyer}) after the substitution $\delta\rightarrow{}-\delta$.

\end{widetext}



\begin{thebibliography}{}

\bibitem{Cas09}
A.H.~Castro Neto, F.~Guinea, N.M.R.~Peres, K.S.~Novoselov, and A.K.~Geim,
Rev.\ Mod.\ Phys.\ \textbf{81}, 109 (2009).

\bibitem{Das11}
S.~Das~Sarma, Sh.~Adam, E.H.~Hwang, and E.~Rossi, 
Rev.\ Mod.\ Phys.\ \textbf{83}, 407 (2011).

\bibitem{Goe11}
M.O.~Goerbig, Rev.\ Mod.\ Phys.\ {\bf 83}, 1193 (2011).

\bibitem{Nov05}
K.S.~Novoselov, A.K.~Geim, S.V.~Morozov, D.~Jiang, M.I.~Katsnelson, I.V.~Grigorieva, S.V.~Dubonos, A.A.~Firsov, and Y.~Zhang, Nature, {\bf 438}, 197 (2005);  Y.~Zhang, Y.-W.~Tan, H.L.~Stormer, and P.~Kim, {\it ibid.} {\bf 438}, 201 (2005).

\bibitem{Nov07}
K.S.~Novoselov, Z.~Jiang, Y.~Zhang, S.V.~Morozov, H.L.~Stormer, U.~Zeitler, J.C.~Maan, G.S.~Boebinger, P.~Kim, and A.K.~Geim, {\bf 315}, 1379 (2007).

\bibitem{Dea13}
C.R.~Dean, L.~Wang, P.~Maher, C.~Forsythe, F.~Ghahari, Y.~Gao, J.~Katoch, M.~Ishigami, P.~Moon, M.~Koshino {\it et al.}, Nature {\bf 497}, 598 (2013); 
L.A.~Ponomarenko, R.V.~Gorbachev, G.L.~Yu, D.C.~Elias, R.~Jalil, A.A.~Patel, A.~Mishchenko, A.S.~Mayorov, C.R.~Woods, J.R.~Wallbank {\it et al.}, {\it ibid.} {\bf 497}, 594 (2013).

\bibitem{linefoo}
In the case of graphene bilayer, the linear nature of the coupling gets unveiled when considering the four-band effective Hamiltonian, see Ref.\ \cite{Mac06}.

\bibitem{Nai08}
R.R.~Nair, P.~Blake, A.N.~Grigorenko, K.S.~Novoselov, T.J.~Booth, T.~Stauber, N.M.R.~Peres, and A.K.~Geim, Science {\bf 320}, 1308 (2008); 
T.~Stauber, N.M.R.~Peres, and A.K.~Geim, Phys.\ Rev.\ B {\bf 78}, 085432 (2008).

\bibitem{Mac06}
E.~McCann, V.I.~Fal'ko, Phys.\ Rev.\ Lett.\ {\bf 96}, 086805 (2006); E.~McCann, 
Phys.\ Rev.\ B {\bf 74}, 161403(R) (2006).

\bibitem{Kat06a}
M.I.~Katsnelson, Eur.\ Phys.\ J.\ B \textbf{51}, 157 (2006).

\bibitem{Two06} 
J.~Tworzyd{\l}o, B.~Trauzettel, M.~Titov, A.~Rycerz, and C.W.J.~Beenakker, 
Phys.\ Rev.\ Lett.\ \textbf{96}, 246802 (2006).

\bibitem{Naz09}
Yu.V.~Nazarov and Ya.M.~Blanter, {\it Quantum Transport: Introduction to Nanoscience,} Cambridge University Press (Cambridge, 2009). 

\bibitem{Ben08}
C.W.J.~Beenakker, Rev.\ Mod.\ Phys.\ \textbf{80}, 1337 (2008).

\bibitem{Ryc09}
For a~generalization of these results for other than rectangular samples, see:
A.~Rycerz, P.~Recher, and M.~Wimmer, Phys.\ Rev.\ B {\bf 80}, 125417 (2009).

\bibitem{Ost06}
P.M.~Ostrovsky, I.V.~Gornyi, A.D.~Mirlin,
Phys.\ Rev.\ B \textbf{74}, 235443 (2006).

\bibitem{Lou07}
E.~Louis,~J. A.~Verges, F.~Guinea, and G.~Chiappe, 
Phys.\ Rev.\ B \textbf{75}, 085440 (2007).

\bibitem{Pra07}
E.~Prada, P.~San-Jose, B.~Wunsch, and F.~Guinea, 
Phys.\ Rev.\ B \textbf{75}, 113407 (2007).

\bibitem{Mia07} 
F.~Miao, S.~Wijeratne, Y.~Zhang, U.C.~Coscun, W.~Bao, and C.N.~Lau, 
Science \textbf{317}, 1530 (2007).

\bibitem{Dic08}
L.~DiCarlo, J.R.~Williams, Y.~Zhang, D.T.~McClure, and C.M.~Marcus,  
Phys.\ Rev.\ Lett.\ {\bf 100}, 156801 (2008).

\bibitem{Dan08}
R.~Danneau, F.~Wu, M.F.~Craciun, S.~Russo, M.Y.~Tomi, J.~Salmilehto, 
A.F.~Morpurgo, and P.J.~Hakonen,
Phys.\ Rev.\ Lett.\ \textbf{100}, 196802 (2008).

\bibitem{Wie11}
A.D.~Wiener and M.~Kindermann, Phys.\ Rev.\ B {\bf 84}, 245420 (2011).

\bibitem{Che06}
V.V.~Cheianov and V.I.~Fal'ko, Phys.\ Rev.\ B \textbf{74}, 041403(R) (2006).

\bibitem{Ryc10}
A.~Rycerz, Phys.\ Rev.\ B {\bf 81}, 121404(R) (2010); M.I.~Katsnelson, Europhys.\ Lett.\ \textbf{89}, 17001 (2010).

\bibitem{Ort13}
F.~Ortmann and S.~Roche, Phys.\ Rev.\ Lett.\ {\bf 110}, 086602 (2013).

\bibitem{Kat06b}
M.I.~Katsnelson, Eur.\ Phys.\ J.\ B {\bf 52}, 151 (2006).

\bibitem{Cse07}
J.~Cserti, Phys.\ Rev.\ B \textbf{75}, 033405 (2007).

\bibitem{Sny07}
I.~Snyman and C.W.J.~Beenakker, Phys.\ Rev.\ B \textbf{75}, 045322 (2007).

\bibitem{Cas07}
E.V.~Castro, K.S.~Novoselov, S.V.~Morozov, N.M.R.~Peres, J.M.B.~Lopes Dos Santos, J.~Nilsson, F.~Guinea, A.K.~Geim, and A.H.~Castro Neto, Phys.\ Rev.\ Lett.\ {\bf 99}, 216802 (2007); J.B.~Oostinga, H.B.~Heersche, X.~Liu, A.F.~Morpurgo, and L.M.K.~Vandersypen, Nat.\ Mater.\ {\bf 7}, 151 (2007).

\bibitem{Nil07}
J.~Nilsson, A.H.~Castro Neto, F.~Guinea, and N.M.R.~Peres, Phys.\ Rev.\ B {\bf 76}, 165416 (2007).

\bibitem{Per07}
J.M.~Pereira, F.~M.~Peeters, and P.~Vasilopoulos, Phys.\ Rev.\ B {\bf 76}, 115419 (2007);
M.~Zarenia, {\it Confined states in mono- and bi-layer graphene nanostructures}, PhD Thesis, Universiteit Antwerpen, Belgium (Antwerpen, 2013). \url{http://www.cmt.ua.ac.be/ua/Zarenia.pdf}

\bibitem{Fer11}
A.~Ferreira, J.~Viana-Gomes, J.~Nilsson, E.R.~Mucciolo, N.M.R.~Peres, and A.H.~Castro Neto, Phys.\ Rev.\ B \textbf{83}, 165402 (2011).

\bibitem{Wei10}
R.T.~Weitz, M.T.~Allen, B.E.~Feldman, J.~Martin, and A.~Yacoby, Science {\bf 330}, 812 (2010).

\bibitem{Lai08}
Y.H.~Lai, J.H.~Ho, C.P.~Chang, and M.F.~Lin, Phys.\ Rev.\ B {\bf 77}, 085426 (2008).

\bibitem{Zha12}
Y.T.~Zhang, X.C.~Xie, and Q.~Sun, Phys.\ Rev.\ B {\bf 86}, 035447 (2012).

\bibitem{Jia07} 
Z.~Jiang, Y.~Zhang, Y.-W.~Tana, H.L.~Stormer, and P.~Kim,
Solid State Communications {\bf 143}, 14 (2007).

\bibitem{Kur11}
E.V.~Kurganova, H.J.~van Elferen, A.~McCollam,
L.A.~Ponomarenko, K.S.~Novoselov, A.~Veligura, B.J.~van Wees, J.C.~Maan, and U.~Zeitler, Phys.\ Rev.\ B {\bf 84}, 121407(R) (2011).

\bibitem{Vol12}
A.V.~Volkov, A.A.~Shylau, and I.V.~Zozoulenko, Phys.\ Rev.\ B {\bf 86}, 155440 (2012).

\bibitem{compexp}
Typically, we took $10^2\Delta-10^3\Delta$ different values of the doping $E$ uniformely distributed over the energy interval $|E-E_0|\leqslant{}{\cal W}_0\Delta/2$, with the scaling factor varied from $\Delta=0.1$ to $100$ with the steps of $0.1$.

\bibitem{Cse07}
J.~Cserti, A.~Csord\'{a}s, and G.~D\'{a}vid, 
Phys.\ Rev.\ Lett.\ {\bf 99}, 066802 (2007); 
M.~Koshino and T.~Ando, Phys.\ Rev.\ B {\bf 73}, 245403 (2006).

\bibitem{Kuz09}
A.B.~Kuzmenko, I.~Crassee, D. van der Marel, P.~Blake and K.S.~Novoselov, 
Phys.\ Rev.\ B {\bf 80}, 165406 (2009).

\bibitem{Mac13}
E.~McCann and M.~Koshino, Rep.\ Prog.\ Phys.\ {\bf 76}, 056503 (2013).

\bibitem{Wan12}
B.~Wang, C.~Zhang, and Z.~Ma, 
J.\ Phys.: Condens.\ Matter {\bf 24}, 485303 (2012); 
{\em ibid.}\ {\bf 24}, 035303 (2012).

\bibitem{May11}
A.S.~Mayorov, D.C.~Elias, M.~Mucha-Kruczy\'{n}ski, R.V.~Gorbachev, T.~Tudorovskiy, A.~Zhukov, S.V.~Morozov, M.I.~Katsnelson, V.I.~Falko, A.K.~Geim, and K.S.~Novoselov, Science {\bf 333}, 860 (2011).

\bibitem{Zie07}
K.~Ziegler, Phys.\ Rev.\ B {\bf 75}, 233407 (2007).

\bibitem{hoppingfoo}
For the numerical calculations, we took $t_0=3.16\,$eV and $t_{\perp}=0.38\,$eV after Ref.\ \cite{Kuz09}. Two different values of $t'=0.16\,$eV and $0.32\,$eV are considered to cover the range of results reported in different works, see Ref.\ \cite{Mac13}.


\bibitem{Olv10}
F.W.J.~Olver, D.W.~Lozier, R.F.~Boisvert, and C.W.~Clark, {\it NIST Handbook of Mathematical Functions}, (Cambridge University Press, Cambridge, 2010), Chapter 16.


\end{thebibliography}
\end{document}